\documentclass[aps,pra,reprint,superscriptaddress,amsmath,amssymb,floatfix]{revtex4-2}
\usepackage{graphicx}
\usepackage{bm}
\usepackage[colorlinks=true,allcolors=blue]{hyperref}
\newcommand{\FV}{\mbox{FV$\frac{1}{2}$}}
\newcommand{\Or}{\mathcal{O}}
\newcommand{\dla}{\langle\!\langle}
\newcommand{\dra}{\rangle\!\rangle}
\newcommand{\eq}[1]{Eq.~(\ref{#1})}
\newcommand{\Eq}[1]{Equation~(\ref{#1})}

\newcommand{\eqsss}[3]{Eqs.~(\ref{#1}), (\ref{#2}), and (\ref{#3})}
\newcommand{\eqsr}[2]{Eqs.~(\ref{#1})--(\ref{#2})}
\newcommand{\sect}[1]{Sec.~\ref{#1}}
\newcommand{\Sect}[1]{Section~\ref{#1}}
\newcommand{\sects}[2]{Secs.~\ref{#1} and \ref{#2}}
\newcommand{\sectsr}[2]{Secs.~\ref{#1}--\ref{#2}}
\newcommand{\Sectsr}[2]{Sections~\ref{#1}--\ref{#2}}
\newcommand{\tabl}[1]{Table~\ref{#1}}
\newcommand{\Tabl}[1]{Table~\ref{#1}}
\newcommand{\tabls}[2]{Tables~\ref{#1} and \ref{#2}}
\newcommand{\fig}[1]{Fig.~\ref{#1}}
\newcommand{\Fig}[1]{Figure~\ref{#1}}
\newcommand{\citeref}[1]{Ref.~\cite{#1}}
\newcommand{\citerefs}[1]{Refs.~\cite{#1}}
\newcommand{\exteq}[1]{Eq.~(#1)}

\begin{document}

\title{Stark effect of hydrogenic ions as a test of the indefinite-metric formalism of the eight-component relativistic wave equation for spin-$\frac12$ particles}

\author{Paulus C. Tjiang}
\email{Corresponding author: pctjiang@unpar.ac.id}
\author{Sylvia H. Sutanto}
\email{sylvia@unpar.ac.id}
\author{Vincentius E. W. Tjia}
\email{vincentiusevanwijaya@gmail.com}
\affiliation{Center for Theoretical Physics, Faculty of Science, Parahyangan Catholic University, Bandung 40141, Indonesia}

\date{\today}

\begin{abstract}
The eight-component relativistic wave equation for spin-$\frac{1}{2}$ particles of Robson and Staudte (FV$\frac{1}{2}$), a Feshbach--Villars linearization of the Feynman--Gell-Mann equation, reproduces the Dirac hydrogenic spectrum with a doubled solution space and an indefinite inner product. Its exact Stark shifts must equal the Dirac ones; whether its indefinite-metric perturbation theory delivers them is not guaranteed. We compute the first- and second-order Stark shifts of the $n=1$--3 levels of hydrogenic ions from H to U$^{91+}$ in the Schr\"odinger, Dirac and FV$\frac{1}{2}$ theories, evaluating all spectral sums, continuum included, by the Dalgarno--Lewis method. The FV$\frac{1}{2}$ and Dirac shifts coincide level by level, to working precision at first order (closed-form roots) and to 45 significant figures at second order. Both ingredients are needed: without the explicit spin--field coupling the first-order roots become complex, and without the negative-norm states the $(Z\alpha)^2$ coefficient is wrong. Norm sign and parity are locked in the doubled space: a potential lifting the $2s_{1/2}$--$2p_{1/2}$ degeneracy needs its own spin--gradient term, and shifts placed by hand on the $\kappa$-labeled states give a spurious imaginary linear Stark shift. The relativistic linear Stark effect is smaller by finite factors, but a second-moment sum rule shows the non-relativistic linear Stark strength is conserved, the missing part reappearing in singular $(Z\alpha)^{-2}$ second-order terms. For levels with a fine-structure partner, an exact intra-shell/regular split of the second-order shift gives the $(Z\alpha)^0$ coefficients analytically and relates the $\pi^2$ terms of the $(Z\alpha)^2$ coefficients to the singular ones. The Dirac weak/strong-field crossover for $n=2$ of hydrogen lies near 2.8~kV cm$^{-1}$, and the perturbative window set by the fine structure grows as $Z^5$.
\end{abstract}

\maketitle

% ===========================================================================
\section{Introduction}\label{sec:intro}

The hydrogenic atom is the reference system against which quantum theories of bound states are
measured. Its spectrum is known in closed form in the Schr\"odinger and Dirac theories, and it is
measured with an accuracy that few other systems can match: the Lamb shift~\cite{Lamb1947} and a series
of recent optical and microwave measurements in atomic
hydrogen~\cite{Beyer2017,Fleurbaey2018,Bezginov2019,Grinin2020,Brandt2022} test bound-state
theory~\cite{Yerokhin2019} at the level that fixes the fundamental constants~\cite{CODATA2022}, while
highly charged hydrogen-like ions probe the same theory in the strong-field regime~\cite{Indelicato2019,Morgner2023}.
The Schr\"odinger theory gives levels that depend on the principal quantum number $n$, the
additional degeneracy in $l$ being a property of the Coulomb potential~\cite{Bethe1957}; the Dirac theory
gives levels that depend on $n$ and on the total angular momentum $j$~\cite{Greiner2000}, and this fine
structure is confirmed experimentally up to the radiative (Lamb-shift) corrections.

A third description is the eight-component relativistic wave equation for spin-$\frac12$ particles of
Robson and Staudte~\cite{RobsonStaudte1996,Staudte1996}. It is obtained from the second-order
(Feynman--Gell-Mann) equation~\cite{FeynmanGellMann1958} by the Feshbach--Villars linearization of the
time derivative~\cite{FeshbachVillars1958}, and we denote it by \FV. It reproduces the Dirac hydrogenic
spectrum exactly, although with different state functions~\cite{RobsonStaudte1996,Staudte1996}. It yields the Klein--Nishina cross section for Compton scattering~\cite{RobsonSutanto2001b}, and radiative transition probabilities that differ from the Dirac ones at large nuclear charge have been reported for it~\cite{RobsonSutanto2001}. Its solution space has twice the dimension of the Dirac one, and its quantum-mechanical formalism rests on the indefinite inner product
\begin{equation}\label{eq:tau5}
\dla\Psi|\Psi\dra=\int\Psi^\dagger(\vec r)\,\tau_5\,\Psi(\vec r)\,{\rm d}^3r,\qquad
\tau_5=\tau_1\otimes\tau_3\otimes\mathbf 1_2,
\end{equation}
with respect to which the \FV\ Hamiltonian is pseudo-Hermitian, $\tau_5\hat H^\dagger\tau_5=\hat H$. States
therefore carry norms of either sign and the closure relation acquires sign weights
$\epsilon_k=\pm1$~\cite{Staudte1996,Bognar1974,Mostafazadeh2002a,Mostafazadeh2010}. Second-order and
pseudo-Hermitian formulations of spin-$\frac12$ dynamics, and Feshbach--Villars formalisms in general,
continue to attract attention~\cite{FerroHernandez2024,Boumali2025}, as does the wider physics of
pseudo-Hermitian and non-Hermitian operators~\cite{Bender2019,ElGanainy2018,MiriAlu2019}.

The Stark effect~\cite{Stark1914}, the shift of the levels in a static uniform electric field, is a
natural probe of such a formalism. Its non-relativistic theory is a classic application of stationary
perturbation theory~\cite{Bethe1957}; its relativistic theory is well developed, from closed-form
polarizabilities based on the Sturmian expansion of the Dirac--Coulomb Green
function~\cite{Szmytkowski1997,SzmytkowskiLukasik2016,Szmytkowski2018} to finite-basis
calculations~\cite{Ivanov2025} and complex-scaling calculations of the Stark energies and widths of hydrogen-like ions~\cite{Maltsev2023}. The relativistic $2s$--$2p$ couplings also provide the Stark input of the field-quenching and anisotropy determinations of the $2s$ Lamb shift~\cite{HilleryMohr1980,GouldMarrus1983,Zacek1984,vanWijngaarden1998,vanWijngaarden2000}. The structure of the problem differs instructively between the
theories. In the Schr\"odinger theory the whole $n=2$ shell is degenerate and the field produces a linear
shift. In the Dirac and \FV\ theories the fine structure splits $n=2$ into a $j=\frac12$ level, spanned by
$2s_{1/2}$ and $2p_{1/2}$, and a $j=\frac32$ level; the residual $2s_{1/2}$--$2p_{1/2}$ degeneracy still
gives a linear shift, whereas the $j=\frac32$ level responds only at second order. The first two orders
therefore exercise both the degenerate and the nondegenerate branches of perturbation theory, and both
the discrete and the continuum parts of the spectrum.

What such a calculation can and cannot test should be stated at the outset. The \FV\ equation is in
one-to-one correspondence with the second-order equation, whose solution space in any external field is
spanned by solutions of the Dirac equations with masses $\pm m$, which have the same
spectrum~\cite{Staudte1996}. The exact \FV\ and Dirac energies in the potential $-Z/r+E_0z$ therefore
coincide, the exact Stark shifts of the two theories are equal, and any discrepancy would signal an error in the perturbative treatment. What is not guaranteed is that the perturbation formalism of the \FV\ theory---indefinite metric, degenerate manifolds of mixed norm sign, biorthogonal left vectors, and an external field that enters twice, through the scalar potential and through an explicit spin--field coupling $\vec\sigma\cdot\vec E$ that has no Dirac counterpart---delivers these shifts order by order, and it is not known in advance which of its ingredients are indispensable. These are the questions addressed here; it is a stronger test of
the formalism than the agreement of the unperturbed spectra, because it involves matrix elements, spectral
sums and norm signs.

The comparison also yields results of independent interest. We obtain all first-order roots for
$n\le3$ in closed form, show by a second-moment sum rule how the linear Stark strength of the
non-relativistic shell survives in the relativistic theories although the relativistic roots are smaller by finite factors, separate every second-order shift exactly into a part internal to the shell and a regular remainder, which fixes much of its $Z\alpha$ expansion analytically, and show how a level shift that lifts the $2s_{1/2}$--$2p_{1/2}$ degeneracy must enter the \FV\ theory for it to supply the couplings used in the field-quenching and anisotropy measurements~\cite{HilleryMohr1980,GouldMarrus1983,Zacek1984,vanWijngaarden1998,vanWijngaarden2000}.

\Sect{sec:pt} collects the perturbation formulas for positive-definite and indefinite inner products. Sections~\ref{sec:schr}, \ref{sec:dirac} and \ref{sec:fv} treat Schr\"odinger, Dirac and \FV\ theories, respectively, for the $n=1$, $2$ and $3$ levels; \sect{sec:fvlevelshift} treats split levels in the \FV\ theory. \Sect{sec:comparison} compares the results: it derives the closed forms and a second-moment sum rule, decomposes the second-order shifts, locates the weak-field/strong-field crossover, delimits the range of validity of the expansion and examines what the Dirac--\FV\ agreement establishes. \Sect{sec:conclusions} concludes. Atomic units ($m_e=e=\hbar=4\pi\varepsilon_0=1$, $c=1/\alpha$) are used throughout; the nucleus is treated as an infinitely heavy point charge $Z$.

% ===========================================================================
\section{Stationary perturbation theory}\label{sec:pt}

Let $\hat H=\hat H_0+\hat H_I$, with $\hat H_0|\psi^{(0)}_n\rangle=E^{(0)}_n|\psi^{(0)}_n\rangle$, and expand
$E_n=\sum_iE^{(i)}_n$ and $|\psi_n\rangle=\sum_i|\psi^{(i)}_n\rangle$ in powers of $\hat H_I$. By Wigner's $2n+1$
rule the first two energy corrections require $|\psi^{(0)}_n\rangle$ and $|\psi^{(1)}_n\rangle$. We assume
that $\hat H_0$ has a real spectrum and a complete set of eigenstates, and that $\hat H_I$ is self-adjoint with
respect to the inner product of the theory.

\subsection{Positive-definite inner product}\label{sec:ptpos}

For the Schr\"odinger and Dirac theories, $\langle\psi^{(0)}_k|\psi^{(0)}_l\rangle=\delta_{kl}$ and
$\sum_k|\psi^{(0)}_k\rangle\langle\psi^{(0)}_k|=1$, the sum running over the whole spectrum. For a
nondegenerate level
\begin{equation}\label{eq:rspt}
\begin{gathered}
E^{(1)}_n=\langle\psi^{(0)}_n|\hat H_I|\psi^{(0)}_n\rangle,\\
E^{(2)}_n=\sum_{k\neq n}\frac{|\langle\psi^{(0)}_k|\hat H_I|\psi^{(0)}_n\rangle|^2}{E^{(0)}_n-E^{(0)}_k},\\
|\psi^{(1)}_n\rangle=\sum_{k\neq n}\frac{\langle\psi^{(0)}_k|\hat H_I|\psi^{(0)}_n\rangle}{E^{(0)}_n-E^{(0)}_k}\,|\psi^{(0)}_k\rangle.
\end{gathered}
\end{equation}
In the Schr\"odinger theory every denominator is negative for the ground state, so that $E^{(2)}_1<0$. In
the Dirac theory this argument does not apply, because the negative-energy continuum enters with positive
denominators; the sign of $E^{(2)}_1$ must then be established by calculation.

If $E^{(0)}_n$ is $f$-fold degenerate, with states $|\psi^{(0)}_{n\alpha}\rangle$, one diagonalizes the
$f\times f$ matrix $W^n_{\alpha\beta}=\langle\psi^{(0)}_{n\alpha}|\hat H_I|\psi^{(0)}_{n\beta}\rangle$,
\begin{equation}\label{eq:secular}
\sum_{\beta=1}^fW^n_{\alpha\beta}\,c^\gamma_\beta=E^{(1)}_{n\gamma}\,c^\gamma_\alpha,\qquad
|\bar\psi^{(0)}_{n\gamma}\rangle=\sum_\beta c^\gamma_\beta|\psi^{(0)}_{n\beta}\rangle.
\end{equation}
The eigenvalues are the first-order corrections and the eigenvectors define the correct zeroth-order
states. In this basis the second-order correction is
\begin{equation}\label{eq:E2deg}
E^{(2)}_{n\gamma}=\sum_{k\notin\{n\}}\frac{|\langle\psi^{(0)}_k|\hat H_I|\bar\psi^{(0)}_{n\gamma}\rangle|^2}{E^{(0)}_n-E^{(0)}_k},
\end{equation}
where $k\notin\{n\}$ excludes the whole degenerate level. If a degeneracy survives the first order, the matrix
\begin{equation}\label{eq:Mmatrix}
M^{(n)}_{\gamma\delta}=\sum_{k\notin\{n\}}\frac{\langle\bar\psi^{(0)}_{n\gamma}|\hat H_I|\psi^{(0)}_k\rangle
\langle\psi^{(0)}_k|\hat H_I|\bar\psi^{(0)}_{n\delta}\rangle}{E^{(0)}_n-E^{(0)}_k}
\end{equation}
must be diagonalized within each surviving subspace; \eq{eq:E2deg} gives its diagonal elements.

\subsection{Indefinite inner product}\label{sec:ptind}

For the \FV\ theory the inner product is \eq{eq:tau5}. Write the $\tau_5$-adjoint as
$\hat A^\ddagger\equiv\tau_5\hat A^\dagger\tau_5$ ($\tau_5=\tau_5^\dagger=\tau_5^{-1}$), so that
$\dla\Phi|\hat A\Psi\dra=\dla\hat A^\ddagger\Phi|\Psi\dra$; an operator is pseudo-Hermitian if $\hat A^\ddagger=\hat A$.
Both $\hat H_0$ and the \FV\ Stark perturbation of \sect{sec:fvpert} have this property. Pseudo-Hermiticity
guarantees neither a real spectrum nor diagonalizability in general~\cite{Mostafazadeh2002a,Mostafazadeh2002b};
for the hydrogenic bound states the spectrum is real~\cite{RobsonStaudte1996}, and, as in \citeref{Staudte1996}, we
assume completeness. The eigenstates are normalized as
\begin{equation}\label{eq:indnorm}
\begin{gathered}
\dla\Psi^{(0)}_k|\Psi^{(0)}_l\dra=\epsilon_k\delta_{kl},\qquad\epsilon_k=\pm1,\\
\sum_k\epsilon_k\,|\Psi^{(0)}_k\rangle\dla\Psi^{(0)}_k|=\mathbf 1_8,
\end{gathered}
\end{equation}
states of negative norm being "maximally normalized" to $-1$~\cite{Staudte1996,Bognar1974}. Expanding
$|\Psi_n\rangle=\sum_kc_k|\Psi^{(0)}_k\rangle$ and projecting gives
\begin{equation}\label{eq:indproj}
\begin{gathered}
\epsilon_m\bigl(E_n-E^{(0)}_m\bigr)c_m=\sum_kW_{mk}c_k,\\
W_{mk}\equiv\dla\Psi^{(0)}_m|\hat H_I|\Psi^{(0)}_k\dra=W^*_{km},
\end{gathered}
\end{equation}
and, with $c_n=1$,
\begin{equation}\label{eq:indrspt}
\begin{gathered}
E^{(1)}_n=\epsilon_nW_{nn},\qquad
E^{(2)}_n=\epsilon_n\sum_{k\neq n}\epsilon_k\frac{|W_{kn}|^2}{E^{(0)}_n-E^{(0)}_k},\\
|\Psi^{(1)}_n\rangle=\sum_{k\neq n}\epsilon_k\frac{W_{kn}}{E^{(0)}_n-E^{(0)}_k}\,|\Psi^{(0)}_k\rangle.
\end{gathered}
\end{equation}
These reduce to \eq{eq:rspt} when all $\epsilon_k=+1$. The weights $\epsilon_k$ appear inside the spectral
sums, so that negative-norm intermediate states enter with the opposite sign, and the enlarged \FV\ solution
space supplies precisely such states.

If $E^{(0)}_n$ is $f$-fold degenerate, with $\dla\Psi^{(0)}_{n\alpha}|\Psi^{(0)}_{n\beta}\dra=\epsilon_{n\alpha}\delta_{\alpha\beta}$,
restricting \eq{eq:indproj} to the manifold gives the generalized secular problem
\begin{equation}\label{eq:gensecular}
\begin{gathered}
W^nc^\gamma=E^{(1)}_{n\gamma}\,\eta^nc^\gamma,\\
W^n_{\alpha\beta}=\dla\Psi^{(0)}_{n\alpha}|\hat H_I|\Psi^{(0)}_{n\beta}\dra,\qquad
\eta^n={\rm diag}(\epsilon_{n1},\dots,\epsilon_{nf}),
\end{gathered}
\end{equation}
with $W^n$ Hermitian and $\eta^n$ Hermitian, nonsingular and in general indefinite; equivalently,
$E^{(1)}_{n\gamma}$ are the eigenvalues of the non-Hermitian matrix $\eta^nW^n$. Eigenvectors belonging to
distinct real roots are $\eta^n$-orthogonal. Roots with non-neutral eigenvectors
($c^{\gamma\dagger}\eta^nc^\gamma\neq0$) are real; as a parameter varies, a pair of real roots typically becomes
complex by colliding at an exceptional point, where the eigenvectors become neutral. Complex-conjugate
pairs always have neutral eigenvectors~\cite{Bognar1974,Mostafazadeh2002a}. Away from such cases a basis
$|\bar\Psi^{(0)}_{n\gamma}\rangle=\sum_\beta c^\gamma_\beta|\Psi^{(0)}_{n\beta}\rangle$ can be chosen with
$\dla\bar\Psi^{(0)}_{n\gamma}|\bar\Psi^{(0)}_{n\delta}\dra=\epsilon_{n\gamma}\delta_{\gamma\delta}$, in which the
perturbation is diagonal, $\dla\bar\Psi^{(0)}_{n\gamma}|\hat H_I|\bar\Psi^{(0)}_{n\delta}\dra=\epsilon_{n\gamma}E^{(1)}_{n\gamma}\delta_{\gamma\delta}$, and
\begin{equation}\label{eq:indE2}
E^{(2)}_{n\gamma}=\epsilon_{n\gamma}\sum_{k\notin\{n\}}\epsilon_k\,
\frac{|\dla\Psi^{(0)}_k|\hat H_I|\bar\Psi^{(0)}_{n\gamma}\dra|^2}{E^{(0)}_n-E^{(0)}_k}.
\end{equation}
Where a degeneracy survives the first order, the matrix
\begin{equation}\label{eq:indM}
M^{(n)}_{\gamma\delta}=\epsilon_{n\gamma}\sum_{k\notin\{n\}}\epsilon_k\,
\frac{\dla\bar\Psi^{(0)}_{n\gamma}|\hat H_I|\Psi^{(0)}_k\dra\dla\Psi^{(0)}_k|\hat H_I|\bar\Psi^{(0)}_{n\delta}\dra}{E^{(0)}_n-E^{(0)}_k}
\end{equation}
must be diagonalized in each surviving subspace; if that subspace carries mixed norm signs the residual problem
is again of the form~\eqref{eq:gensecular}. Only when a manifold carries a single norm sign does
\eq{eq:gensecular} reduce to the Hermitian problem~\eqref{eq:secular}; \sect{sec:fvdoubled} shows that this never
happens for the \FV\ hydrogenic atom.

\subsection{The Stark perturbation}\label{sec:units}

For a static, uniform field along $z$, $\vec E=E_0\hat e_z$, the electron acquires the potential energy $E_0 z$, and the perturbation Hamiltonian is given by
\begin{equation}\label{eq:HI}
\hat H_I=E_0\,r\cos\theta.
\end{equation}
\Eq{eq:HI} is in the Schr\"odinger and Dirac theories, in which a
static external field enters the Hamiltonian only through the scalar potential. In the \FV\ theory, the field also enters through the spin term of the Feynman--Gell-Mann equation, which survives the Feshbach--Villars linearization as an explicit $\vec\sigma\cdot\vec E$ coupling (\sect{sec:fvpert}). Three remarks fix the regime.
(i)~$\hat H_I$ is odd under parity; in the Schr\"odinger and Dirac theories it is spin independent and conserves
$m$ (respectively $m_j$). (ii)~Because $\hat H_I$ is unbounded below, the bound states become resonances whose
widths are nonperturbative in $E_0$~\cite{Maltsev2023}; the perturbation series is asymptotic, and for the fields
considered here the widths are negligible. (iii)~The expansion requires $|E^{(2)}|\ll|E^{(1)}|$ whenever
$E^{(1)}\neq0$. In the relativistic theories the second order is enhanced by $(Z\alpha)^{-2}$ whenever the level
has a fine-structure partner, which makes this condition restrictive for light ions (\sect{sec:validity}).

% ===========================================================================
\section{Stark effect in the nonrelativistic hydrogenic atom}\label{sec:schr}

The unperturbed problem is $\hat H_0=-\frac12\nabla^2-Z/r$, with $E^{(0)}_n=-Z^2/(2n^2)$ and, in terms of the spin-angular functions needed later,
\begin{equation}\label{eq:schrstates}
\psi^{(0)}_{nljm_j}(\vec r)=R_{nl}(r)\sum_{m,m_s}C\bigl(l\tfrac12j;m\,m_s\,m_j\bigr)\,Y_{lm}\,\chi_{\frac12m_s},
\end{equation}
where $j=l\pm\frac12$, the Clebsch--Gordan coefficients follow the Condon--Shortley convention and $R_{nl}$ is
the usual radial function, positive near the origin~\cite{Bethe1957}. Since $\hat H_I$ is spin independent,
the correct zeroth-order states can always be chosen as (orbital state)$\otimes$(spin state), and every root
is twofold degenerate in $m_s$. The level $E^{(0)}_n$ is $2n^2$-fold degenerate, so the whole shell enters
the secular problem~\eqref{eq:secular}.

\emph{Scaling in $Z$.} Under $r\to r/Z$ the unperturbed problem maps onto the $Z=1$ problem with energies
multiplied by $Z^2$, while $\hat H_I$ acquires a factor $1/Z$. Hence
\begin{equation}\label{eq:scaling}
E^{(1)}=f_1E_0/Z,\qquad E^{(2)}=f_2E_0^2/Z^4,
\end{equation}
with pure numbers $f_1$, $f_2$. In \sects{sec:dirac}{sec:fv} these become functions of $Z\alpha$.

\emph{Second order.} We do not evaluate \eq{eq:E2deg} as a sum over states. Instead, the Dalgarno--Lewis
equation~\cite{DalgarnoLewis1955,Mossman2016,AntaoPeres2021}
\begin{equation}\label{eq:DL}
\begin{gathered}
\bigl(\hat H_0-E^{(0)}_n\bigr)|\psi^{(1)}\rangle=-\bigl(\hat H_I-E^{(1)}\bigr)|\bar\psi^{(0)}\rangle,\\
E^{(2)}=\langle\bar\psi^{(0)}|\hat H_I|\psi^{(1)}\rangle,
\end{gathered}
\end{equation}
with $|\psi^{(1)}\rangle=\sum_{l'}u_{l'}(r)Y_{l'm}\chi_{\frac12m_s}$ orthogonal to the level $n$, separates into one
radial equation per channel $l'$,
\begin{equation}\label{eq:DLradial}
\begin{split}
\biggl[-\frac12\biggl(\frac{{\rm d}^2}{{\rm d}r^2}+\frac2r\frac{\rm d}{{\rm d}r}-\frac{l'(l'+1)}{r^2}\biggr)
-\frac Zr-E^{(0)}_n\biggr]u_{l'}\qquad&\\
=-E_0\,r\sum_la_l\,c_{l'lm}R_{nl}+E^{(1)}a_{l'}R_{nl'},&
\end{split}
\end{equation}
where $a_l$ are the amplitudes of the correct zeroth-order state and
$c_{l+1,l,m}=c_{l,l+1,m}=[((l+1)^2-m^2)/((2l+1)(2l+3))]^{1/2}$ are the matrix elements of $\cos\theta$. The
ansatz $u_{l'}={\rm e}^{-Zr/n}\sum_kb_kr^k$ terminates, so the solution is exact and includes the full continuum contribution.

\subsection{Energy correction for $n=1$}\label{sec:schr1}

The $n=1$ level ($E^{(0)}_1=-Z^2/2$) is spanned by $\psi_{10\frac12\pm\frac12}=(Z^3/\pi)^{1/2}{\rm e}^{-Zr}\chi_{\frac12\pm\frac12}$.
Both states are even, so $W^1=0$ and $E^{(1)}=0$. The single open channel $l'=1$ has the Dalgarno--Lewis solution
\begin{equation}\label{eq:schr1s}
u_1(r)=-\frac{E_0\,r(2+Zr)}{\sqrt3\sqrt Z}\,{\rm e}^{-Zr},\qquad E^{(2)}=-\frac{9}{4Z^4}E_0^2,
\end{equation}
i.e., the exact static dipole polarizability $\alpha_d=9/(2Z^4)$.

\begin{table*}[tbp]
\caption{Stark corrections of the Schr\"odinger hydrogenic atom, $n=1,2,3$. Here $(k,m)$ are the parabolic
quantum numbers and $g$ is the degeneracy including spin; the entries are independent of $Z$. Columns 6--12 give the sum
over the bound intermediate states $\beta$ of principal quantum number $n_\beta\le n_{\max}$ (outside the shell),
$\sum_\beta|\langle\beta|\hat H_I|\bar\psi^{(0)}\rangle|^2/(E^{(0)}_n-E^{(0)}_\beta)$, in units of $E_0^2/Z^4$; the last column is the $n_{\max}=100$ sum as a percentage of the exact value.}
\label{tab:schr}
\begin{ruledtabular}
\begin{tabular}{cccrrrrrrrrrr}
$n$ & $(k,m)$ & $g$ & $ZE^{(1)}/E_0$ & $Z^4E^{(2)}/E_0^2$ & \multicolumn{7}{c}{bound-state sum, $n_{\max}=$} & \\
\cline{6-12}
 &  &  &  & (exact) & 3 & 4 & 6 & 10 & 20 & 50 & 100 & \% \\
\hline
1 & $(0,0)$ & 2 & $0$ & $-9/4$ & $-1.6800$ & $-1.7460$ & $-1.7928$ & $-1.8172$ & $-1.8279$ & $-1.8310$ & $-1.8315$ & 81.4 \\
2 & $(\pm1,0)$ & 4 & $\pm3$ & $-84$ & $-67.20$ & $-74.54$ & $-77.78$ & $-79.03$ & $-79.49$ & $-79.62$ & $-79.64$ & 94.8 \\
2 & $(0,\pm1)$ & 4 & $0$ & $-78$ & $-64.93$ & $-71.16$ & $-73.76$ & $-74.72$ & $-75.07$ & $-75.17$ & $-75.18$ & 96.4 \\
3 & $(\pm2,0)$ & 4 & $\pm9$ & $-810$ & $+47.50$ & $-673.0$ & $-764.9$ & $-783.4$ & $-788.7$ & $-790.0$ & $-790.2$ & 97.6 \\
3 & $(\pm1,\pm1)$ & 8 & $\pm\frac92$ & $-810$ & $+32.46$ & $-702.1$ & $-779.3$ & $-793.1$ & $-796.9$ & $-797.8$ & $-797.9$ & 98.5 \\
3 & $(0,0)$ & 2 & $0$ & $-3483/4$ & $+41.09$ & $-763.9$ & $-841.5$ & $-854.8$ & $-858.4$ & $-859.2$ & $-859.3$ & 98.7 \\
3 & $(0,\pm2)$ & 4 & $0$ & $-1377/2$ & $0$ & $-615.1$ & $-670.5$ & $-679.4$ & $-681.6$ & $-682.2$ & $-682.2$ & 99.1 \\
\end{tabular}
\end{ruledtabular}
\end{table*}

\subsection{Energy correction for $n=2$}\label{sec:schr2}

For $n=2$ ($E^{(0)}_2=-Z^2/8$) the eight-dimensional secular matrix, written in the coupled
basis~\eqref{eq:schrstates}, has as its only nonvanishing elements those of
\begin{equation}\label{eq:schr2me}
\begin{gathered}
\langle2s_{1/2}m_j|\hat H_I|2p_{1/2}m_j\rangle=\pm\frac{\sqrt3E_0}{Z}\qquad\bigl(m_j=\pm\tfrac12\bigr),\\
\langle2s_{1/2}m_j|\hat H_I|2p_{3/2}m_j\rangle=-\frac{\sqrt6E_0}{Z},
\end{gathered}
\end{equation}
and their transposes. Its eigenvalues are $E^{(1)}=\{\pm3\,(\times2),\,0\,(\times4)\}\,E_0/Z$, since
$\sqrt{3+6}=3$. The correct zeroth-order states are the parabolic states
\begin{equation}\label{eq:schr2states}
\begin{gathered}
\bar\psi_\pm=\tfrac{1}{\sqrt2}\bigl(|200\rangle\mp|210\rangle\bigr)\otimes\chi_{\frac12m_s},\qquad E^{(1)}=\pm\frac{3E_0}{Z};\\
\bar\psi_0=|21{\pm}1\rangle\otimes\chi_{\frac12m_s},\qquad E^{(1)}=0,
\end{gathered}
\end{equation}
with $k\equiv n_1-n_2=\pm1$, $m=0$ and $k=0$, $|m|=1$. The first order resolves the degeneracy in $k$, and $m$ is conserved, so \eq{eq:E2deg} applies directly and gives $Z^4E^{(2)}/E_0^2=-84$ for $(k,m)=(\pm1,0)$ and $-78$ for $(0,\pm1)$.

\subsection{Energy correction for $n=3$}\label{sec:schr3}

The eighteen-dimensional $n=3$ secular problem has the roots
$E^{(1)}=\{\pm9\,(\times2),\pm\frac92\,(\times4),0\,(\times6)\}\,E_0/Z$, in agreement with
$E^{(1)}=\frac32nkE_0/Z$ for $k=0,\pm1,\pm2$. All exact second-order corrections, collected in \tabl{tab:schr},
reproduce the parabolic formula~\cite{Bethe1957}
\begin{equation}\label{eq:parabolic}
E^{(2)}=-\frac{n^4}{16Z^4}\bigl[17n^2-3k^2-9m^2+19\bigr]E_0^2
\end{equation}
exactly, which is the first check of the machinery used in \sects{sec:dirac}{sec:fv}.

\subsection{The role of the continuum}\label{sec:schrcont}

\Tabl{tab:schr} compares the exact corrections with the sum over bound intermediate states truncated at
$n_{\max}$. The truncated sum converges, but to the total bound-state contribution, which is smaller in modulus
than the exact value; the difference is the continuum contribution, which no increase of $n_{\max}$ can
recover~\cite{Ruffa1973}. For the ground state the whole discrete spectrum supplies $81.4\%$ of $\alpha_d$,
the continuum supplying the remaining $18.6\%$. The deficit decreases with $n$, because the dominant intermediate
states then lie closer to the perturbed level, but it remains at the percent level at $n=3$. At $n_{\max}=3$ the
sums for $n=3$ have the wrong sign or vanish, because only the \emph{lower-lying} $n=1$ and $2$ levels, which contribute
positively, are then included. A second-order Stark calculation restricted to bound intermediate states would
therefore be quantitatively wrong, and the Dalgarno--Lewis construction is used throughout.

% ===========================================================================
\section{Relativistic hydrogenic atom: the Dirac theory}\label{sec:dirac}

The stationary Dirac equation $\{c\vec\alpha\cdot\vec p+\beta c^2-Z/r\}\psi=E\psi$ has the bound-state eigenvalues
\begin{equation}\label{eq:diracE}
E_{n,j}=c^2\Biggl[1+\frac{Z^2\alpha^2}{\Bigl(n-j-\frac12+\sqrt{(j+\frac12)^2-Z^2\alpha^2}\Bigr)^2}\Biggr]^{-1/2}
\end{equation}
(including the rest energy) and the state functions
\begin{equation}\label{eq:diracstates}
\begin{gathered}
\psi_{njlm_j}=\begin{pmatrix}g(r)\,\Omega_{\kappa m_j}\\ {\rm i}f(r)\,\Omega_{-\kappa m_j}\end{pmatrix},\\
\Omega_{\kappa m_j}=\sum_{m,m_s}C\bigl(l\tfrac12j;m\,m_s\,m_j\bigr)Y_{lm}\chi_{\frac12m_s},
\end{gathered}
\end{equation}
with $\kappa=-(l+1)$ for $j=l+\frac12$ and $\kappa=+l$ for $j=l-\frac12$; $\Omega_{-\kappa m_j}$ carries $\bar l=2j-l$, and
$\vec\sigma\cdot\hat r\,\Omega_{\kappa m_j}=-\Omega_{-\kappa m_j}$. With $G=rg$, $F=rf$ and $V=-Z/r$,
\begin{equation}\label{eq:diracradial}
\begin{gathered}
\frac{{\rm d}G}{{\rm d}r}+\frac\kappa rG=\frac{E-V+c^2}{c}\,F,\\
\frac{{\rm d}F}{{\rm d}r}-\frac\kappa rF=-\frac{E-V-c^2}{c}\,G,
\end{gathered}
\end{equation}
and $\int_0^\infty(G^2+F^2){\rm d}r=1$. With $\gamma=\sqrt{\kappa^2-Z^2\alpha^2}$, $\tilde n=n-|\kappa|$,
$N=\sqrt{\tilde n^2+2\tilde n\gamma+\kappa^2}$ and $\varepsilon\equiv E_{n,j}/c^2=(\tilde n+\gamma)/N$, the radial functions
are~\cite{Greiner2000,RobsonSutanto2001}
\begin{widetext}
\begin{equation}\label{eq:diracfg}
\begin{split}
\begin{Bmatrix}g(r)\\ f(r)\end{Bmatrix}=&\pm\frac{[\Gamma(2\gamma+\tilde n+1)]^{1/2}}{\Gamma(2\gamma+1)(\tilde n!)^{1/2}}
\biggl[\frac{1\pm\varepsilon}{4N(N-\kappa)}\biggr]^{1/2}\biggl(\frac{2Z}{N}\biggr)^{3/2}\biggl(\frac{2Zr}{N}\biggr)^{\gamma-1}{\rm e}^{-Zr/N}\\
&\times\biggl[\mp\tilde n\,F\Bigl(1-\tilde n,2\gamma+1;\frac{2Zr}{N}\Bigr)+(N-\kappa)\,F\Bigl(-\tilde n,2\gamma+1;\frac{2Zr}{N}\Bigr)\biggr],
\end{split}
\end{equation}
\end{widetext}
with the upper (lower) signs for $g$ ($f$); the relative sign of $g$ and $f$ is the one required by \eq{eq:diracradial}. The
normalization and \eq{eq:diracradial} were verified numerically for every state used below.

The Dirac levels depend on $(n,j)$ but not on $l$, so the exactly degenerate manifolds are labeled by $(n,j)$ and contain at most
the two values $l=j\mp\frac12$. At fixed $m_j$ the perturbation~\eqref{eq:HI} therefore acts in a manifold of dimension one or two,
instead of the $2n^2$-dimensional shell of \sect{sec:schr}. This fact is the origin of the qualitative difference between
the results of this section and those of \sect{sec:schr}. The matrix elements are
\begin{equation}\label{eq:diracme}
\begin{split}
\langle\psi_{n'j'l'm_j}|\hat H_I|\psi_{njlm_j}\rangle={}&E_0A_{\kappa'\kappa}\int_0^\infty G'G\,r\,{\rm d}r\\
&+E_0A_{-\kappa',-\kappa}\int_0^\infty F'F\,r\,{\rm d}r,\\
A_{\kappa'\kappa}\equiv{}&\langle\Omega_{\kappa'm_j}|\cos\theta|\Omega_{\kappa m_j}\rangle,
\end{split}
\end{equation}
and, since $G$ and $F$ are polynomials times $r^\gamma{\rm e}^{-Zr/N}$, the radial integrals are combinations of $\Gamma$ functions, so the results are exact in $Z\alpha$.

For the second order the Dalgarno--Lewis equation~\eqref{eq:DL} becomes, in each channel $\kappa'$, the inhomogeneous version
of \eq{eq:diracradial} with the projections of $-(\hat H_I-E^{(1)})|\bar\psi^{(0)}\rangle$ as sources; the channels reached from a
level built on $|\kappa|$ have $|\kappa'|\in\{|\kappa|-1,|\kappa|,|\kappa|+1\}$ and $j'\ge|m_j|$. We solve these equations by a Galerkin projection. In each channel the large and the small component are expanded in the same functions
\begin{equation}\label{eq:basis}
u_p(r)=r^{\gamma'+p}\,{\rm e}^{-\Lambda r},\qquad p=0,\dots,P,\qquad\Lambda=\frac ZN,
\end{equation}
with one restriction: the lowest pair enters in the combination $(G,F)=(u_0,\rho\,u_0)$, with $\rho=c(\gamma'+\kappa')/Z$, the
ratio $F/G$ of the regular solutions of \eq{eq:diracradial} at $r\to0$. Three features were found to be necessary. (i)~The basis
has the correct index $\gamma'$ at the origin and the decay constant $\Lambda$ of the perturbed level, so that all matrix elements are
$\Gamma$ functions. (ii)~Whenever the channel and source indices differ by a noninteger (every channel with $|\kappa'|\neq|\kappa|$),
a second family $r^{\gamma+1+p}{\rm e}^{-\Lambda r}$ built on the source index is adjoined, without which the convergence in $P$ is algebraic. (iii)~The lowest large- and small-component functions are tied together as described ("origin balance"), so that the basis is compatible with the exact small-$r$ behavior. With independent expansions of $G$ and $F$, the Galerkin matrix of the channel $\kappa'=+|\kappa|$ of a nodeless level ($1s_{1/2}$, $2p_{3/2}$, $3d_{5/2}$) has an exact spurious null vector, $F=\tilde\rho\,G$ with $\tilde\rho=c(\kappa'-\gamma')/Z=1/\rho$, the ratio $F/G$ of the irregular solutions: a function with the regular power $r^{\gamma'}$
but the component ratio of the irregular solution. Tying the lowest pair removes it and enforces the correct small-$r$ behavior. The kinetically balanced small-component basis $v_p=({\rm d}/{\rm d}r+\kappa'/r)\,u_p$, standard for bound Dirac
states~\cite{StantonHavriliak1984} (see also the dual kinetic balance of \citeref{Shabaev2004}), is not suitable here: for $\kappa'>0$
every $v_p$ satisfies $\int_0^\infty r^{\kappa'}v_p\,{\rm d}r=0$, a constraint that the exact Dalgarno--Lewis small component violates, so the expansion converges algebraically. For the ground state of U$^{91+}$ it gives $Z^4E^{(2)}/E_0^2=-1.252\,845$,
$-1.252\,806$ and $-1.252\,799$ at $P=8$, $14$ and $20$, compared to $-1.252\,794$; for the $(n,j,|m_j|)=(2,\frac32,\frac12)$ level of U$^{91+}$ the error is large and decreases slowly with $P$ ($828.2$, $631.8$ and $551.0$ at the same $P$,
compared to the converged $153.0479$), whereas the origin-balanced basis gives $153.047\,890\,755\,656\,18$ already at $P=8$. Members of the
degenerate manifold are projected out by a bordered linear system. The Galerkin matrices are ill-conditioned, so the arithmetic is
carried out with 130--160 decimal digits. No second-order value reported below changes by more than $10^{-36}$ (relative) between
$P=16$ and $P=22$, and in four test cases, including the $1s_{1/2}$ and $(2,\frac32,\frac12)$ levels of U$^{91+}$, the change from $P=21$ to
$P=22$ is below $3\times10^{-48}$. This rapid convergence relies on the decay constant $\Lambda=Z/N$: with $\Lambda$ changed by $\pm40\%$ the
results at $P=22$ are accurate to about $2\times10^{-17}$ (relative).

Once lengths are measured in units of $1/Z$ and energies in units of $Z^2$, the Dirac--Coulomb problem depends on $Z$ and $\alpha$ only through $Z\alpha$. The relations~\eqref{eq:scaling} therefore remain valid in the form
\begin{equation}\label{eq:universal}
ZE^{(1)}/E_0=f_1(Z\alpha),\qquad Z^4E^{(2)}/E_0^2=f_2(Z\alpha),
\end{equation}
which we verified numerically, and \tabls{tab:table2rev}{tab:ions} tabulate these two universal functions. For the levels without a
fine-structure partner their values at $Z\alpha=0$ are the nonrelativistic results; for the others the limit $Z\alpha\to0$ is
discussed in \sect{sec:comparison}.

\subsection{Energy correction for $n=1$}\label{sec:dirac1}

The level $n=1$, $j=\frac12$ contains only $l=0$, so $E^{(1)}_{1\frac12\pm\frac12}=0$, and
\begin{equation}\label{eq:dirac1s}
E^{(2)}_{1\frac12\pm\frac12}=-\frac{9}{4Z^4}\Bigl[1-\frac{28}{27}(Z\alpha)^2+\Or\bigl((Z\alpha)^4\bigr)\Bigr]E_0^2,
\end{equation}
i.e., $\alpha_d=\frac{9}{2Z^4}[1-\frac{28}{27}(Z\alpha)^2+\dots]$. The leading term is \eq{eq:schr1s} and the coefficient $-28/27$
is the known relativistic correction to the dipole polarizability of the Dirac one-electron
atom~\cite{Szmytkowski1997,SzmytkowskiLukasik2016}; recovering it as an output of the Dalgarno--Lewis solution is the second check of
the machinery. Numerically, with $\alpha=1/137$,
\begin{equation}\label{eq:alphad}
\begin{split}
Z^4\alpha_d=\{&4.499\,751,\ 4.499\,005,\ 4.497\,762,\\ &4.332\,609,\ 3.787\,855,\ 2.505\,588\}
\end{split}
\end{equation}
for H, He$^+$, Li$^{2+}$, Fe$^{25+}$, Xe$^{53+}$ and U$^{91+}$: the relativistic contraction of the ground state reduces the scaled
polarizability by $0.006\%$ in hydrogen and by $44\%$ in U$^{91+}$. Like the other levels without a fine-structure partner
(\sect{sec:regular}), the $n=1$ level has relativistic and nonrelativistic second-order shifts of the same order of magnitude.

\subsection{Energy correction for $n=2$}\label{sec:dirac2}

The eight states of $n=2$ form the fourfold $j=\frac12$ level ($2s_{1/2}$, $2p_{1/2}$) and the fourfold $j=\frac32$ level
($2p_{3/2}$), separated by
\begin{equation}\label{eq:Delta2}
\Delta_2\equiv E_{2,\frac32}-E_{2,\frac12}=\frac{Z^4\alpha^2}{32}+\Or(Z^6\alpha^4).
\end{equation}
For $j=\frac12$ and each $m_j$, $W^2$ is a $2\times2$ matrix, purely off-diagonal by parity, whose roots are
\begin{equation}\label{eq:f1n2}
E^{(1)}_{2\frac12}=\pm\frac{\sqrt3E_0}{Z}\Bigl[1-\frac{5}{12}(Z\alpha)^2-\frac{11}{144}(Z\alpha)^4+\dots\Bigr],
\end{equation}
with the equal-weight combinations of $|2s_{1/2}m_j\rangle$ and $|2p_{1/2}m_j\rangle$ as the correct zeroth-order states. This is not a small correction to the nonrelativistic value  $\pm3E_0/Z$: the linear shift is smaller by the finite factor $\sqrt3$, because $2p_{3/2}$, which supplies the element $\sqrt6E_0/Z$ of \eq{eq:schr2me}, is no longer degenerate with $2s_{1/2}$. The $j=\frac32$ level is not shifted at first order, $E^{(1)}_{2\frac32m_j}=0$, since $\hat H_I$ has no matrix element within a manifold of a single $l$.

At second order the fine-structure partner re-enters through an energy denominator of order $Z^4\alpha^2$ instead of $Z^2$. We write the second-order
shifts of the states $(n,j,|m_j|)$ as $Z^4E^{(2)}/E_0^2=c_{-2}/(Z\alpha)^2+c_0+c_2(Z\alpha)^2+\dots$. For $|m_j|=\frac12$ both levels have a partner and
\begin{equation}\label{eq:E2n2}
\begin{gathered}
E^{(2)}_{2\frac12\pm\frac12}=\frac{E_0^2}{Z^4}\Bigl[-\frac{96}{(Z\alpha)^2}+18+\Or\bigl((Z\alpha)^2\bigr)\Bigr],\\
E^{(2)}_{2\frac32\pm\frac12}=\frac{E_0^2}{Z^4}\Bigl[\frac{192}{(Z\alpha)^2}-282+\Or\bigl((Z\alpha)^2\bigr)\Bigr],
\end{gathered}
\end{equation}
whereas for $|m_j|=\frac32$ there is no partner and
$E^{(2)}_{2\frac32\pm\frac32}=Z^{-4}[-78+\frac{93}{4}(Z\alpha)^2+\Or((Z\alpha)^4)]E_0^2$. The $(Z\alpha)^2$ terms of \eq{eq:E2n2}
are given in \tabl{tab:table2rev}.

\subsection{Energy correction for $n=3$}\label{sec:dirac3}

The eighteen states of $n=3$ form the levels $j=\frac12$ ($3s_{1/2}$, $3p_{1/2}$), $j=\frac32$ ($3p_{3/2}$, $3d_{3/2}$) and $j=\frac52$
($3d_{5/2}$). The first two are two-dimensional at each $m_j$ and are split linearly; the third is one-dimensional and and has no linear shift. All results for $n=1$, $2$ and $3$ are collected in \tabl{tab:table2rev} as expansions in $Z\alpha$ and in \tabl{tab:ions} as numbers for
six ions.

\begin{table*}[tbp]
\caption{Stark corrections of the Dirac hydrogenic atom, which coincide with those of the \FV\ theory (\sect{sec:fv}). The entries are the universal functions~\eqref{eq:universal}, $x\equiv Z\alpha$, $\gamma_k\equiv(k^2-x^2)^{1/2}$. The first-order roots are the closed forms~\eqref{eq:f1closed}; their expansions are shown below them. For the second order, $Z^4E^{(2)}/E_0^2=c_{-2}/x^2+c_0+c_2x^2+\dots$. The coefficients $c_{-2}$ and $c_0$ as well as the coefficients of $\pi^2$ in the $x^2$ terms, are known analytically (for the levels without a fine-structure partner $c_0$ is the Schr\"odinger value; for the others see \sect{sec:partner}). The remaining coefficients were extracted by polynomial interpolation in $x^2$ of $x^2f_2$ through ten values at $x=i/1000$ ($i=1,\dots,10$). They are stable against an eight-point interpolation to at least 26 significant figures for the $x^2$ and 23 for the $x^4$ coefficients, and were identified as rationals or, by an integer-relation search, as $a+b\pi^2$; the analytically known coefficients agree with the interpolated ones, and the $x^4$ coefficient of the ground state is confirmed by the closed form of Szmytkowski~\cite{Szmytkowski1997}. Each root occurs for both signs of $m_j$; for the two-dimensional manifolds both roots of $E^{(1)}$ are shown, and they share the same $E^{(2)}$.}
\label{tab:table2rev}
\begin{ruledtabular}
\begin{tabular}{ccll}
$n$ & $(j,|m_j|)$ & $ZE^{(1)}/E_0$ & $Z^4E^{(2)}/E_0^2$ \\
\hline
1 & $(\frac12,\frac12)$ & $0$ & $-\frac94+\frac73x^2-\bigl(\frac{31}{192}+\frac{\pi^2}{96}\bigr)x^4+\Or(x^6)$ \\[3pt]
2 & $(\frac12,\frac12)$ & \begin{tabular}[t]{@{}l@{}}$\pm\frac12(1+\gamma_1)\sqrt{1+2\gamma_1}$\\[1pt] $=\pm\sqrt3\,[1-\frac{5}{12}x^2-\frac{11}{144}x^4-\dots]$\end{tabular} & $-\frac{96}{x^2}+18+\bigl(\frac{533}{6}+\pi^2\bigr)x^2+\Or(x^4)$ \\[9pt]
2 & $(\frac32,\frac12)$ & $0$ & $+\frac{192}{x^2}-282+\bigl(\frac{587}{12}-2\pi^2\bigr)x^2+\Or(x^4)$ \\[3pt]
2 & $(\frac32,\frac32)$ & $0$ & $-78+\frac{93}{4}x^2-\bigl(\frac{85}{576}+\frac{\pi^2}{12}\bigr)x^4+\Or(x^6)$ \\[3pt]
3 & $(\frac12,\frac12)$ & \begin{tabular}[t]{@{}l@{}}$\pm(2+\gamma_1)\sqrt{1+\gamma_1}$\\[1pt] $=\pm3\sqrt2\,[1-\frac{7}{24}x^2-\frac{23}{384}x^4-\dots]$\end{tabular} & $-\frac{3159}{x^2}+\frac{34425}{16}+\bigl(\frac{11643}{20}+\frac{1053}{32}\pi^2\bigr)x^2+\Or(x^4)$ \\[9pt]
3 & $(\frac32,\frac32)$ & \begin{tabular}[t]{@{}l@{}}$\pm\frac{3}{10}(1+\gamma_2)\sqrt{1+2\gamma_2}$\\[1pt] $=\pm\frac{9}{2\sqrt5}\,[1-\frac{2}{15}x^2-\frac{13}{2400}x^4-\dots]$\end{tabular} & $-\frac{13122}{5x^2}+\frac{10773}{200}+\bigl(\frac{451557}{2000}+\frac{243}{80}\pi^2\bigr)x^2+\Or(x^4)$ \\[9pt]
3 & $(\frac32,\frac12)$ & \begin{tabular}[t]{@{}l@{}}$\pm\frac{1}{10}(1+\gamma_2)\sqrt{1+2\gamma_2}$\\[1pt] $=\pm\frac{3}{2\sqrt5}\,[1-\frac{2}{15}x^2-\frac{13}{2400}x^4-\dots]$\end{tabular} & $-\frac{3888}{5x^2}-\frac{256689}{100}+\bigl(\frac{986223}{2000}-\frac{567}{20}\pi^2\bigr)x^2+\Or(x^4)$ \\[9pt]
3 & $(\frac52,\frac52)$ & $0$ & $-\frac{1377}{2}+\frac{1359}{14}x^2+\bigl(\frac{32549}{44800}-\frac{9}{32}\pi^2\bigr)x^4+\Or(x^6)$ \\[3pt]
3 & $(\frac52,\frac32)$ & $0$ & $+\frac{26244}{5x^2}-\frac{241623}{100}+\bigl(\frac{194472}{875}-\frac{243}{40}\pi^2\bigr)x^2+\Or(x^4)$ \\[3pt]
3 & $(\frac52,\frac12)$ & $0$ & $+\frac{39366}{5x^2}-\frac{656019}{200}+\bigl(\frac{996957}{3500}-\frac{729}{80}\pi^2\bigr)x^2+\Or(x^4)$ \\[3pt]
\end{tabular}
\end{ruledtabular}
\end{table*}

\begin{table*}[tbp]
\caption{Stark corrections of the Dirac (and \FV) hydrogenic atom for six ions, $\alpha=1/137$.
Upper line of each entry: $ZE^{(1)}/E_0$ (the second root is its negative); lower line: $Z^4E^{(2)}/E_0^2$. The Dirac and \FV\ values
agree to all 45 significant figures retained. The notation $a(b)$ means $a\times10^b$.}
\label{tab:ions}
\begin{ruledtabular}
\begin{tabular}{ccrrrrrr}
$n$ & $(j,|m_j|)$ & H & He$^{+}$ & Li$^{2+}$ & Fe$^{25+}$ & Xe$^{53+}$ & U$^{91+}$ \\
\hline
 & $Z\alpha=$ & $0.00730$ & $0.01460$ & $0.02190$ & $0.18978$ & $0.39416$ & $0.67153$ \\
\hline
1 & $(\frac12,\frac12)$ & $0$ & $0$ & $0$ & $0$ & $0$ & $0$ \\
 &  & $-2.249876$ & $-2.249503$ & $-2.248881$ & $-2.166304$ & $-1.893927$ & $-1.252794$ \\
2 & $(\frac12,\frac12)$ & $1.732012$ & $1.731897$ & $1.731705$ & $1.705883$ & $1.616467$ & $1.371384$ \\
 &  & $-1.801806(6)$ & $-4.504380(5)$ & $-2.001846(5)$ & $-2.643911(3)$ & $-5.854190(2)$ & $-1.574021(2)$ \\
2 & $(\frac32,\frac12)$ & $0$ & $0$ & $0$ & $0$ & $0$ & $0$ \\
 &  & $+3.603366(6)$ & $+9.006300(5)$ & $+4.001233(5)$ & $+5.049874(3)$ & $+9.580017(2)$ & $+1.530479(2)$ \\
2 & $(\frac32,\frac32)$ & $0$ & $0$ & $0$ & $0$ & $0$ & $0$ \\
 &  & $-77.99876$ & $-77.99505$ & $-77.98885$ & $-77.16387$ & $-74.41134$ & $-67.71514$ \\
3 & $(\frac12,\frac12)$ & $4.242575$ & $4.242377$ & $4.242047$ & $4.197736$ & $4.043733$ & $3.616607$ \\
 &  & $-5.928912(7)$ & $-1.482067(7)$ & $-6.585767(6)$ & $-8.552517(4)$ & $-1.804822(4)$ & $-4.506628(3)$ \\
3 & $(\frac32,\frac32)$ & $2.012447$ & $2.012404$ & $2.012333$ & $2.002783$ & $1.970505$ & $1.889114$ \\
 &  & $-4.925731(7)$ & $-1.231429(7)$ & $-5.472986(6)$ & $-7.280289(4)$ & $-1.679923(4)$ & $-5.656591(3)$ \\
3 & $(\frac32,\frac12)$ & $0.6708156$ & $0.6708013$ & $0.6707775$ & $0.6675942$ & $0.6568350$ & $0.6297047$ \\
 &  & $-1.459734(7)$ & $-3.651260(6)$ & $-1.624208(6)$ & $-2.414936(4)$ & $-7.543943(3)$ & $-4.252438(3)$ \\
3 & $(\frac52,\frac52)$ & $0$ & $0$ & $0$ & $0$ & $0$ & $0$ \\
 &  & $-688.4948$ & $-688.4793$ & $-688.4535$ & $-685.0065$ & $-673.4683$ & $-645.1445$ \\
3 & $(\frac52,\frac32)$ & $0$ & $0$ & $0$ & $0$ & $0$ & $0$ \\
 &  & $+9.851231(7)$ & $+2.462627(7)$ & $+1.094366(7)$ & $+1.433214(5)$ & $+3.139298(4)$ & $+9.294439(3)$ \\
3 & $(\frac52,\frac12)$ & $0$ & $0$ & $0$ & $0$ & $0$ & $0$ \\
 &  & $+1.477688(8)$ & $+3.693974(7)$ & $+1.641584(7)$ & $+2.153247(5)$ & $+4.742621(4)$ & $+1.426423(4)$ \\
\end{tabular}
\end{ruledtabular}
\end{table*}

Two exact sum rules connect the singular and constant coefficients of \tabl{tab:table2rev} with the nonrelativistic results
(\sect{sec:partner} explains them through a decomposition of the second-order shift). Because
two levels of a shell shift each other by equal and opposite amounts, the $(Z\alpha)^{-2}$ coefficients cancel at fixed $(n,|m_j|)$:
\begin{equation}\label{eq:sumrule1}
\begin{gathered}
2(-96)+192=0,\qquad2\Bigl(-\frac{13122}{5}\Bigr)+\frac{26244}{5}=0,\\
2(-3159)+2\Bigl(-\frac{3888}{5}\Bigr)+\frac{39366}{5}=0
\end{gathered}
\end{equation}
for $n=2$, $|m_j|=\frac12$, for $n=3$, $|m_j|=\frac32$, and for $n=3$, $|m_j|=\frac12$. The same cancellation removes every intra-shell
contribution from the trace, at every order in $Z\alpha$, so the trace of $Z^4E^{(2)}/E_0^2$ over a shell must reduce at $Z\alpha=0$ to
the nonrelativistic trace of \tabl{tab:schr}. Both shells satisfy this exactly:
\begin{equation}\label{eq:sumrule2}
\begin{gathered}
4(18)+2(-282)+2(-78)=-648=4(-84)+4(-78),\\
4\,\frac{34425}{16}+4\,\frac{10773}{200}-4\,\frac{256689}{100}\\
-2\,\frac{1377}{2}-2\,\frac{241623}{100}-2\,\frac{656019}{200}=-\frac{28431}{2}\\
=12(-810)+2\Bigl(-\frac{3483}{4}\Bigr)+4\Bigl(-\frac{1377}{2}\Bigr).
\end{gathered}
\end{equation}

% ===========================================================================
\section{Relativistic hydrogenic atom: the \FV\ theory}\label{sec:fv}

In the Weyl representation the \FV\ equation of Robson and Staudte~\cite{RobsonStaudte1996} decouples into two four-component
equations, ${\rm i}\partial_t\Psi_\xi=\hat H_\xi\Psi_\xi$ and ${\rm i}\partial_t\Psi_{\dot\eta}=\hat H_{\dot\eta}\Psi_{\dot\eta}$,
with~\cite{RobsonStaudte1996,Staudte1996,RobsonSutanto2001}
\begin{equation}\label{eq:fvH}
\begin{split}
\hat H_{\xi,\dot\eta}={}&(\tau_3+{\rm i}\tau_2)\otimes\frac12\Bigl[-\mathbf D^2\pm{\rm i}\alpha q\,\vec\sigma\cdot\bigl(\vec E\pm{\rm i}\vec B\bigr)\Bigr]\\
&+c^2\,\tau_3\otimes\mathbf 1_2+qA_0\,\mathbf 1_4,
\end{split}
\end{equation}
in atomic units, where $q$ is the charge of the particle ($q=-1$ for the electron), $\mathbf D$ the minimally coupled gradient,
$\tau_3+{\rm i}\tau_2$ and $c^2\tau_3$ the Feshbach--Villars structures~\cite{FeshbachVillars1958}, and the upper (lower) signs refer
to $\xi$ ($\dot\eta$). For a static electric field, with potential energy $V=qA_0$ and $q\vec E=-\nabla V$,
\begin{equation}\label{eq:fvHstatic}
\begin{split}
\hat H_{\xi,\dot\eta}={}&(\tau_3+{\rm i}\tau_2)\otimes\Bigl[\frac{p^2}{2}\mp\frac{{\rm i}\alpha}{2}\,\vec\sigma\cdot\nabla V\Bigr]\\
&+c^2\,\tau_3\otimes\mathbf 1_2+V\,\mathbf 1_4,
\end{split}
\end{equation}
which for $V=-Z/r$ contains the spin--orbit-generating term $\mp\frac{\rm i}{2}Z\alpha\,(\tau_3+{\rm i}\tau_2)\otimes\vec\sigma\cdot\hat r/r^2$
of \citerefs{RobsonStaudte1996,RobsonSutanto2001}. The full eight-component inner product is \eq{eq:tau5}; since $\tau_3(\hat H_{\dot\eta})^\dagger\tau_3=\hat H_\xi$, the Hamiltonian ${\rm diag}(\hat H_\xi,\hat H_{\dot\eta})$ is $\tau_5$-pseudo-Hermitian.

With $\kappa$, $\gamma$, $\tilde n$, $N$ and $\varepsilon$ as in \sect{sec:dirac}, define
\begin{equation}\label{eq:fvparams}
\begin{gathered}
\bar\gamma=\begin{cases}\gamma-1,&j=l+\frac12\\ \gamma,&j=l-\frac12\end{cases},\\
Z\alpha\,\bar\kappa=\begin{cases}\kappa+\gamma,&j=l+\frac12\\ \kappa-\gamma,&j=l-\frac12\end{cases},\\
\tilde n'=\tilde n+\gamma-\bar\gamma,\qquad\Lambda=\frac ZN,
\end{gathered}
\end{equation}
so that $\bar\kappa=\mp Z\alpha/(|\kappa|+\gamma)$ for $j=l\pm\frac12$ and $\tilde n'+\bar\gamma=\tilde n+\gamma=\varepsilon N$; the last
relation terminates the confluent hypergeometric series below and fixes the spectrum ($\tilde n'\ge1$ always, whereas the Dirac $\tilde n$
starts at 0). The bound states are~\cite{RobsonStaudte1996,Staudte1996,RobsonSutanto2001}
\begin{equation}\label{eq:fvstates}
\begin{gathered}
\Psi_\xi=\begin{pmatrix}\bar g\,\Theta_\xi\\ \bar f\,\Theta_\xi\end{pmatrix},\qquad
\Psi_{\dot\eta}=\begin{pmatrix}\bar g\,\Theta_{\dot\eta}\\ \bar f\,\Theta_{\dot\eta}\end{pmatrix},\\
\Theta_{\xi,\dot\eta}=\Omega_{\kappa m_j}\pm{\rm i}\bar\kappa\,\Omega_{-\kappa m_j},
\end{gathered}
\end{equation}
\begin{equation}\label{eq:fvradial}
\begin{gathered}
\begin{Bmatrix}\bar g(r)\\ \bar f(r)\end{Bmatrix}=\mathcal N(2\Lambda r)^{\bar\gamma}{\rm e}^{-\Lambda r}F\bigl(1-\tilde n',2\bar\gamma+2;2\Lambda r\bigr)\\
\times\Bigl\{1\pm\Bigl(\varepsilon+\frac{\alpha^2Z}{r}\Bigr)\Bigr\},\\
\mathcal N=\frac{\Lambda^2[\Gamma(2\bar\gamma+\tilde n'+1)]^{1/2}}{\Gamma(2\bar\gamma+2)\,[Z(\tilde n'-1)!]^{1/2}\,[2(1-\bar\kappa^2)]^{1/2}},
\end{gathered}
\end{equation}
and the eigenvalues are the Dirac ones~\eqref{eq:diracE}. It is convenient to use
\begin{equation}\label{eq:fvsd}
\begin{gathered}
s=\bar g+\bar f\propto r^{\bar\gamma}{\rm e}^{-\Lambda r}F\bigl(1-\tilde n',2\bar\gamma+2;2\Lambda r\bigr),\\
d=\bar g-\bar f=\alpha^2\Bigl(E+\frac Zr\Bigr)s,
\end{gathered}
\end{equation}
where $E=c^2\varepsilon$; the second relation is an exact consequence of the radial \FV\ equations, and $s$ obeys the Coulomb-like
equation $-s''-\frac2rs'+\frac{\bar\gamma(\bar\gamma+1)}{r^2}s-\frac{2\varepsilon Z}{r}s+\frac{Z^2}{N^2}s=0$. The functions $\bar g$ and
$\bar f$ separately behave as $r^{\bar\gamma-1}$ at the origin (for the $ns_{1/2}$ states, where $\bar\gamma=\gamma-1<0$, they diverge slightly faster than $1/r$); it is the combination $\bar g^2-\bar f^2=sd$, whose density $sd\,r^2$ vanishes like $r^{2\gamma-1}$ in the worst case, that
the $\tau_5$ inner product requires. The normalization,
\begin{equation}\label{eq:fvnorm}
\dla\Psi|\Psi\dra=2\bigl(1-\bar\kappa^2\bigr)\int_0^\infty s\,d\,r^2\,{\rm d}r=1,
\end{equation}
reproduces the constants $C$ and $C_0$ of \citeref{Staudte1996}. We verified that \eqsr{eq:fvstates}{eq:fvsd} satisfy the radial
equations, are $\tau_5$-normalized, are $\tau_5$-orthogonal for different $n$ in the same channel and reproduce \eq{eq:diracE}, for
every state used below and for $Z\alpha$ up to $0.6715$.

\subsection{The Stark perturbation in the \FV\ theory}\label{sec:fvpert}

With $V=-Z/r+E_0z$ in \eq{eq:fvHstatic}, the Coulomb part reproduces the unperturbed Hamiltonian and the field contributes
\begin{equation}\label{eq:fvHI}
\begin{gathered}
\hat H_I^\xi=E_0\,r\cos\theta\,\mathbf 1_4-\frac{{\rm i}\alpha E_0}{2}(\tau_3+{\rm i}\tau_2)\otimes\sigma_z,\\
\hat H_I^{\dot\eta}=E_0\,r\cos\theta\,\mathbf 1_4+\frac{{\rm i}\alpha E_0}{2}(\tau_3+{\rm i}\tau_2)\otimes\sigma_z.
\end{gathered}
\end{equation}
The eight-component perturbation $\hat H_I={\rm diag}(\hat H_I^\xi,\hat H_I^{\dot\eta})$ satisfies $\tau_5\hat H_I^\dagger\tau_5=\hat H_I$, as
\sect{sec:ptind} requires, although neither block is $\tau_3$-pseudo-Hermitian on its own. The spin term is not an electric dipole
moment of the electron: it is the external-field part of the same spin--field coupling that, for the Coulomb field, generates the fine
structure of the \FV\ spectrum.

This spin term cannot be neglected. Because $\bar\kappa=\Or(Z\alpha)$, the parity mixing in $\Theta_\xi$ lets the scalar term connect states of equal $l$, and both this contribution and the spin term (which carries an explicit $\alpha$) enter the matrix elements at relative order
$Z\alpha$. The role of the spin term is clearest at $n=1$, where the manifold is one-dimensional in each block and the first-order shift is $2Y_{11}$ in the notation of \eq{eq:Ymatrix} below. The scalar term alone gives
\begin{equation}\label{eq:Y11scalar}
2Y_{11}\big|_{\rm scalar}=\frac{\rm i}{2}Z\alpha\,\frac{E_0}{Z}\bigl[1+\Or\bigl((Z\alpha)^2\bigr)\bigr],
\end{equation}
which is purely imaginary, so that the secular problem would have the complex-conjugate roots $\pm\frac{\rm i}{2}\alpha E_0$---precisely the possibility that
\sect{sec:ptind} leaves open for a manifold of mixed norm sign. The spin term cancels this contribution identically: for every $Z\alpha$ we
examined, up to $0.6715$, $Y_{11}$ vanishes to the full working precision and $E^{(1)}_{1s}=0$ is recovered.

\subsection{The doubled solution space and the secular problem}\label{sec:fvdoubled}

Because ${\rm diag}(\hat H_\xi,\hat H_{\dot\eta})$ commutes with ${\rm diag}(\mathbf 1_4,-\mathbf 1_4)$, both $(\Psi_\xi,\Psi_{\dot\eta})$ and
$(\Psi_\xi,-\Psi_{\dot\eta})$ are eigenstates with the same energy---the $\delta=0$ and $\delta=\pi$ solutions of \citeref{Staudte1996}---and since
$\tau_5$ is off-diagonal in the $(\xi,\dot\eta)$ index their norms are opposite,
\begin{equation}\label{eq:fvpm}
|\Psi^{(\pm)}\rangle\equiv(\Psi_\xi,\pm\Psi_{\dot\eta}),\qquad\dla\Psi^{(\pm)}|\Psi^{(\pm)}\dra=\pm1.
\end{equation}
Every bound level is therefore doubly degenerate with both norm signs present, and the manifold at $(n,j,m_j)$, spanned by the
pairs~\eqref{eq:fvpm} with $l=j\mp\frac12$, has dimension twice the number of $l$ values (2 or 4) and mixed norm sign. Writing
\begin{equation}\label{eq:XY}
\begin{gathered}
X_{\alpha\beta}=\int\Psi^\dagger_{\xi\alpha}\tau_3\hat H_I^{\dot\eta}\Psi_{\dot\eta\beta}\,{\rm d}^3r,\\
Y_{\alpha\beta}=\int\Psi^\dagger_{\dot\eta\alpha}\tau_3\hat H_I^\xi\Psi_{\xi\beta}\,{\rm d}^3r,\qquad X=Y^\dagger,
\end{gathered}
\end{equation}
one has $\dla\Psi^{(\epsilon)}_\alpha|\hat H_I|\Psi^{(\epsilon')}_\beta\dra=\epsilon'X_{\alpha\beta}+\epsilon Y_{\alpha\beta}$, so that the
quantities of \eq{eq:gensecular} are
\begin{equation}\label{eq:fvW}
\begin{gathered}
W^n=\begin{pmatrix}X+Y&Y-X\\ X-Y&-(X+Y)\end{pmatrix},\qquad
\eta^n=\begin{pmatrix}\mathbf 1&0\\0&-\mathbf 1\end{pmatrix},\\
\eta^nW^n=\begin{pmatrix}X+Y&Y-X\\ Y-X&X+Y\end{pmatrix}.
\end{gathered}
\end{equation}
The first-order corrections are the eigenvalues of $2Y$ together with those of $2X=(2Y)^\dagger$, i.e., the eigenvalues of $2Y$ and their
complex conjugates. The partners mix whenever $X\neq Y$, which is already the case for the scalar term of \eq{eq:fvHI} alone, because
$\Theta_\xi\neq\Theta_{\dot\eta}$ [\eq{eq:Y11scalar}]. Restricting the secular problem to the positive-norm states $\Psi^{(+)}$, i.e., treating the manifold as if it carried a single norm sign, would instead give the eigenvalues of $X+Y$, a different matrix.

In practice the block structure is exploited directly: $\hat H_0$ and $\hat H_I$ are simultaneously block diagonal, and it suffices to work
in the $\xi$ block. In this block $\hat H_\xi$ is not Hermitian, but $\tau_3\hat H_{\dot\eta}^\dagger\tau_3=\hat H_\xi$, so its left eigenvectors are
supplied by the $\dot\eta$ block, $\langle L_k|=2\langle\Psi_{\dot\eta k}|\tau_3$ with $\langle L_k|\Psi_{\xi l}\rangle=\delta_{kl}$ for $\tau_5$-normalized states, and \sect{sec:ptind} reduces to biorthogonal perturbation theory in that block. The first-order shifts are the
eigenvalues of $2Y$, and its right and left eigenvectors define the correct zeroth-order states. Each real root appears twice, once from each block. The two eigenvectors of $\eta^nW^n$ belonging to one real root, $(u,u)$ and $(w,-w)$ with $2Yu=\lambda u$ and $2Xw=\lambda w$,
correspond, through \eq{eq:fvpm}, to the purely $\xi$ and purely $\dot\eta$ combinations $(\Psi_\xi,0)$ and $(0,\Psi_{\dot\eta})$. Each has zero norm, while their mutual pairing $\int\Psi_\xi^\dagger\tau_3\Psi_{\dot\eta}$ does not vanish, so the metric on the eigenspace has signature $(+,-)$, the problem is not at an exceptional point, and the $\pm1$-normalized basis of \sect{sec:ptind} exists.

For two states of the same $m_j$,
\begin{equation}\label{eq:Ymatrix}
\begin{split}
Y_{BK}={}&\frac{E_0}{2}R^z_{BK}\langle\Theta_{\dot\eta B}|\cos\theta|\Theta_{\xi K}\rangle\\
&-\frac{{\rm i}\alpha E_0}{2}R^s_{BK}\langle\Theta_{\dot\eta B}|\sigma_z|\Theta_{\xi K}\rangle,\\
R^z_{BK}={}&\int_0^\infty(s_Bd_K+d_Bs_K)\,r^3{\rm d}r,\\
R^s_{BK}={}&\int_0^\infty s_Bs_K\,r^2{\rm d}r,
\end{split}
\end{equation}
with, for example,
$\langle\Theta_{\dot\eta B}|\cos\theta|\Theta_{\xi K}\rangle=A_{\kappa_B\kappa_K}+{\rm i}\bar\kappa_KA_{\kappa_B,-\kappa_K}+{\rm i}\bar\kappa_BA_{-\kappa_B,\kappa_K}
-\bar\kappa_B\bar\kappa_KA_{-\kappa_B,-\kappa_K}$, and similarly for $\sigma_z$. All radial integrals are again $\Gamma$ functions. For every
level and every ion considered, the roots of $2Y$ are real and coincide with the Dirac roots of \sect{sec:dirac}.

\subsection{Second-order corrections}\label{sec:fvsecond}

Following \eq{eq:indE2}, with the sum over the two $\tau_5$ partners of each intermediate level (the negative-norm partner entering with
$\epsilon_k=-1$) already carried out, the second order becomes a biorthogonal Dalgarno--Lewis problem in the $\xi$ block,
\begin{equation}\label{eq:fvDL}
\begin{gathered}
\bigl(\hat H_\xi-E\bigr)|\Psi^{(1)}_\xi\rangle=-\bigl(\hat H_I^\xi-E^{(1)}\bigr)|\bar\Psi^{(0)}_\xi\rangle,\\
E^{(2)}=\langle\bar L|\hat H_I^\xi|\Psi^{(1)}_\xi\rangle,
\end{gathered}
\end{equation}
where $E$ is the unperturbed energy and $\langle\bar L|$ is the left partner of $|\bar\Psi^{(0)}_\xi\rangle$. In the variables~\eqref{eq:fvsd} the four-component radial problem in
a channel $(l',j')$ collapses to one scalar equation. With the source projected on that channel written as $(\rho_s,\rho_d)$, the relation
$d=\alpha^2(E+Z/r)s+\alpha^2\rho_s$ eliminates $d$ and leaves
\begin{equation}\label{eq:fvscalar}
\begin{split}
-s''-\frac2rs'+\frac{\bar\gamma'(\bar\gamma'+1)}{r^2}s-\frac{2\varepsilon Z}{r}s+\frac{Z^2}{N^2}s\qquad&\\
=\rho_d+\alpha^2\Bigl(E+\frac Zr\Bigr)\rho_s,&
\end{split}
\end{equation}
a Coulomb-like equation with effective angular momentum $\bar\gamma'$, effective charge $\varepsilon Z$ and binding parameter $Z/N$. It is solved with the Galerkin construction of \sect{sec:dirac} using the single basis $r^{\bar\gamma'+p}{\rm e}^{-\Lambda r}$ (no balance condition is needed for a scalar equation), supplemented by the second exponent family when $\bar\gamma'-\bar\gamma$ is not an integer. Two points are essential. First, the channels reached by $\hat H_I$ are all $(l',j')$ with $|j'-j|\le1$ and $l'=j'\mp\frac12$---twice as many as in the Dirac case, because $\Theta_\xi$ mixes $l$ and $\bar l$. Second, both the parity-mixed scalar term and the $\sigma_z$ term connect a state to its own channel. Omitting either set of additional channel couplings changes $E^{(2)}$ at relative order $(Z\alpha)^2$.

\subsection{Results and comparison with the Dirac theory}\label{sec:fvresults}

The \FV\ first- and second-order corrections for $n=1$, $2$, $3$ and all six ions are those listed in \tabls{tab:table2rev}{tab:ions}. At first order the \FV\ and Dirac roots, both obtained from exact $\Gamma$-function matrix elements, agree to all 40 or more digits retained at every level and every $Z$, including $Z\alpha=0.6715$, which indicates that the agreement is exact in $Z\alpha$. At second order the two calculations, carried out
through separate code paths (the origin-balanced two-component radial system of \sect{sec:dirac} for Dirac, the scalar equation~\eqref{eq:fvscalar} with biorthogonal left vectors for \FV), are limited in precision by the Galerkin solution and agree to all 45 significant figures retained. The imaginary parts of the \FV\ shifts, whose vanishing is not enforced in the implementation, are below $10^{-49}$.

This agreement is what the equivalence of the \FV\ and Dirac eigenvalue problems (\sect{sec:intro}) requires, and it shows that the indefinite-metric perturbation theory of \sect{sec:ptind}, applied with the perturbation~\eqref{eq:fvHI} and the full doubled manifolds,
delivers these shifts order by order. It is not a statement that the state functions are the same: they are not, and \sect{sec:DvsFV} shows which parts of the spectral sums the two theories share and which they do not.

\begin{table}[tbp]
\caption{First-order root of the $n=2$, $j=\frac12$ level of the \FV\
theory, $ZE^{(1)}/(\sqrt3E_0)$, with and without the $\vec\sigma\cdot\vec E$ term of \eq{eq:fvHI}, and with the full mixed-norm secular
problem~\eqref{eq:fvW} or with the positive-norm states $\Psi^{(+)}$ only [roots of $X+Y$]; $x=Z\alpha$, terms of higher order omitted.
The Dirac value is $1-\frac{5}{12}x^2+\Or(x^4)$.}
\label{tab:ingredients}
\begin{ruledtabular}
\begin{tabular}{lll}
 & with $\vec\sigma\cdot\vec E$ & without $\vec\sigma\cdot\vec E$ \\
\hline
full manifold & $1-\frac{5}{12}x^2$ (Dirac) & $1-\frac{47}{108}x^2+\frac{{\rm i}\,x}{6\sqrt3}$ \\[3pt]
$\Psi^{(+)}$ only & $1-\frac{1}{24}x^2$ & $1+\frac18x^2$ \\[3pt]
\end{tabular}
\end{ruledtabular}
\end{table}

\subsection{Which ingredients are indispensable}\label{sec:fvingredients}

Each of the two ingredients examined, the $\vec{\sigma} \cdot \vec{E}$ term and the negative-norm partners, is necessary, as \tabl{tab:ingredients} shows for the upper first-order root $ZE^{(1)}/(\sqrt3E_0)$ of the $n=2$, $j=\frac12$ level. Only the complete indefinite-metric secular problem with the spin--field term retained returns the Dirac value. Without the spin--field term the roots are not even real: every level acquires an imaginary first-order shift of order $\alpha E_0$. For the roots of $2Y$ one finds, to leading order, ${\rm Im}\,ZE^{(1)}/E_0=\pm Z\alpha\,m_j/(2l+1)$ for a one-dimensional manifold with $j=l\pm\frac12$, and the mean of the two corresponding values for a two-dimensional manifold; the roots of $2X$ are the complex conjugates. Restricting the manifold to the positive-norm partners gives real roots, but with the wrong $(Z\alpha)^2$ coefficient. Either omission would make the \FV\ theory appear to disagree with the Dirac theory at, or below, the order of the fine structure that it reproduces.

\subsection{Split levels: how a level shift enters the \FV\ theory}\label{sec:fvlevelshift}

In real ions the $2s_{1/2}$--$2p_{1/2}$ degeneracy of the Coulomb problem is lifted by the finite size of the nucleus
and by the radiative corrections, and in weak fields the two levels are shifted at second order through their
coupling $v_1E_0=\langle2s_{1/2}|\hat H_I|2p_{1/2}\rangle$ of \eq{eq:diracme}. This coupling and the
$2s_{1/2}$--$2p_{3/2}$ coupling $v_3E_0$ are the relativistic Stark input of the field-quenching and anisotropy
determinations of the $2s$ Lamb shift~\cite{HilleryMohr1980,GouldMarrus1983,Zacek1984,vanWijngaarden1998,vanWijngaarden2000},
in which the level energies are taken from experiment or from bound-state QED; \sect{sec:first} compares $v_1$ and $v_3$ with the
values used there. In the Dirac theory
the level-shift operator is diagonal in $\{2s_{1/2},2p_{1/2},2p_{3/2}\}$, because it conserves $j$, $m_j$ and
parity~\cite{HilleryMohr1980}, and the shifts can simply be added to the diagonal of the Stark problem. In the \FV\ theory
this step requires care, because the four-component states $\Psi_\xi$ and $\Psi_{\dot\eta}$ of \eq{eq:fvstates}, and hence the $\kappa$ labeled states, are not parity eigenstates.

Parity acts on the eight-component states as $P:(\Psi_\xi(\vec r),\Psi_{\dot\eta}(\vec r))\mapsto(\Psi_{\dot\eta}(-\vec r),
\Psi_\xi(-\vec r))$, which exchanges $\hat H_\xi$ and $\hat H_{\dot\eta}$ for a central potential and commutes with $\tau_5$. Since
$\Theta_\xi(-\hat r)=(-1)^l\Theta_{\dot\eta}(\hat r)$, the partners~\eqref{eq:fvpm} satisfy
\begin{equation}\label{eq:fvparity}
P\,\Psi^{(\pm)}=\pm(-1)^l\,\Psi^{(\pm)}.
\end{equation}
Norm sign and parity are therefore locked: the two norm partners of a state have opposite parity. The Stark operator is odd;
besides coupling $\Psi^{(+)}$ of $\kappa$ to $\Psi^{(+)}$ of $-\kappa$, it couples the two partners of the same $\kappa$,
which in the $\xi$ block appears as imaginary diagonal elements of $2Y$. For $n=2$, $j=m_j=\frac12$,
\begin{equation}\label{eq:Yn2}
2Y=\begin{pmatrix}{\rm i}A&b\\ b&-{\rm i}A\end{pmatrix},\qquad A=\tfrac32\alpha E_0\bigl[1+\Or\bigl((Z\alpha)^2\bigr)\bigr],
\end{equation}
in the basis of the $\xi$-block states $|\kappa=-1\rangle$ and $|\kappa=+1\rangle$ (the $\Psi_\xi$ of \eq{eq:fvstates}), with $b$ real; $b=\sqrt3[1-(Z\alpha)^2/24+\dots]E_0/Z$ is the positive-norm root of
\tabl{tab:ingredients}, and the roots $\pm(b^2-A^2)^{1/2}=\pm|v_1|E_0$ are the Dirac ones.

A change $\delta V(r)$ of the electrostatic potential energy, due to, e.g., the finite nuclear size or vacuum polarization in the Uehling approximation~\cite{Yerokhin2019}, enters \eq{eq:fvHstatic} exactly as the Stark potential does,
\begin{equation}\label{eq:fvdV}
\hat H_\delta^\xi=\delta V\,\mathbf 1_4-\frac{{\rm i}\alpha}{2}\,\delta V'(r)\,(\tau_3+{\rm i}\tau_2)\otimes\vec\sigma\cdot\hat r.
\end{equation}
The resulting perturbation is even: within a manifold it couples, besides each state to itself, $\Psi^{(+)}$ of $\kappa$ to $\Psi^{(-)}$ of $-\kappa$. In the $\xi$ block its scalar
part is diagonal, because $\bar\kappa$ changes sign with $\kappa$ [\eq{eq:fvparams}], so the coupling of the $\kappa=-1$ and
$\kappa=+1$ states comes entirely from the spin--gradient term. We tested \eq{eq:fvdV} for $\delta V=\lambda{\rm e}^{-\mu Zr}$
($\mu=1$, $4$, $20$) and $\delta V=\lambda{\rm e}^{-4Zr}/(Zr)$ on the $j=\frac12$ manifolds of $n=2$ and $3$ and the $j=\frac32$
manifold of $n=3$, for the six ions: the eigenvalues of the matrix of \eq{eq:fvdV} agree with the Dirac expectation values $\langle\delta V\rangle$ to at least 37 significant figures, and all matrix elements forbidden by \eq{eq:fvparity} vanish. For a uniformly charged nuclei of radii $0.01/Z$ and $0.1/Z$ (atomic units), the agreement at $n=2$ is limited by the quadrature to at least 23 figures. Without the spin--gradient term the shifts are in error by relative amounts ranging from $7\times10^{-6}$ (H, longest range) to more than $100\%$ (U$^{91+}$, shortest range).

With the field and $\delta V$ together, the eigenvalues on the $n=2$, $j=\frac12$ manifold coincide with those of the Dirac matrix with $\Delta_{s,p}=\langle\delta V\rangle_{2s,2p}$ on the diagonal and $v_1E_0$ off it, to working precision, for couplings from $10^{-3}$ to $10^2$ times the splitting, i.e., from the quadratic to the linear regime. If instead $\Delta_s$ and $\Delta_p$ are placed on the diagonal of the $\kappa=-1$ and $\kappa=+1$ states, as in the Dirac treatment, the $2s$ level becomes $\Delta_s+{\rm i}A+b^2/(\Delta_s-\Delta_p)+\dots$ instead of
$\Delta_s+(b^2-A^2)/(\Delta_s-\Delta_p)+\dots$: it acquires an imaginary linear shift $\frac32{\rm i}\alpha E_0$, independent of $Z$ and for the two signs of $m_j$, and its quadratic coefficient becomes that of the positive-norm coupling. The linear term is an artifact of the naive insertion, but not a small one: it would change the decay rates of the two $m_j$
components by $\mp3\alpha E_0$ (atomic units). In hydrogen at $100$~V\,cm$^{-1}$ this amounts to $1.8\times10^7$~s$^{-1}$, comparable with the quench rate itself ($2.8\times10^7$~s$^{-1}$), which is quadratic in the field (\exteq{29} of \citeref{HilleryMohr1980}).

The correct zeroth-order states of the split levels are therefore not the $\kappa$-labeled states but the combinations
$t_s=|\kappa=-1\rangle+c\,|\kappa=+1\rangle$ and $t_p=-c\,|\kappa=-1\rangle+|\kappa=+1\rangle$ in which $2Y$ takes the Dirac form,
with off-diagonal elements $|v_1|E_0$ and zero diagonal. From \eq{eq:Yn2},
\begin{equation}\label{eq:cmix}
c=-{\rm i}\,\frac{b-|v_1|E_0}{A}=-{\rm i}\,\frac{\sqrt3}{4}\,Z\alpha\bigl[1+\Or\bigl((Z\alpha)^2\bigr)\bigr],
\end{equation}
and the matrix of \eq{eq:fvdV} is diagonal in $\{t_s,t_p\}$ for every $\delta V$ tested, the finite nucleus included, to working precision. The same basis serves for the $2p_{3/2}$ level: the Stark matrix elements between $2p_{3/2}$ and $t_p$ vanish, and the product of those between $2p_{3/2}$ and $t_p$ is $(v_3E_0)^2$, whereas in the $\kappa$ basis the $\kappa=+1$ state also couples to $2p_{3/2}$. Since $t_s$ and $t_p$ are fixed by the Stark matrix alone and do not depend on $\delta V$, level shifts of any origin, taken from experiment or computed in QED, should be placed on the diagonal in $\{t_s,t_p\}$; the \FV\ theory then yields the Dirac couplings on which the quench analyses rest.

\section{Comparison of the three theories}\label{sec:comparison}

The Dirac and \FV\ shifts coincide at every level and every order computed
(\sect{sec:fvresults}), so this section compares two descriptions (the
nonrelativistic one and the relativistic one shared by the Dirac and \FV\
theories) and returns in \sect{sec:DvsFV} to what the coincidence
establishes about the \FV\ formalism. Two results are added to those of
\sectsr{sec:schr}{sec:fv}: closed forms of all first-order roots,
and an exact decomposition of every second-order shift into a part internal to
the shell and a part due to all other states. Together they determine part of
\tabl{tab:table2rev} analytically: the first-order roots, the singular
coefficients $c_{-2}$, the constant terms $c_0$ and the coefficients of $\pi^2$
in $c_2$ are derived, while the rational parts of $c_2$ and the $(Z\alpha)^4$
coefficients of the levels without a fine-structure partner are identified from
the numerical results (caption of \tabl{tab:table2rev}). Sum rules connect
the relativistic coefficients to the nonrelativistic ones.

\subsection{Why the theories differ: the degenerate manifolds}\label{sec:manifolds}

The $2n^2$-fold degeneracy of the Schr\"odinger shell reflects the conservation
of the Runge--Lenz vector in the Coulomb field~\cite{Bethe1957}; the electric field mixes
all values of $l$ in the shell, and the parabolic quantum numbers diagonalize the
first-order problem. Relativity removes this symmetry. What survives in the Dirac-Coulomb problem is the degeneracy of $\kappa$ and $-\kappa$ at fixed $n$ and $j$,
associated with the Johnson--Lippmann operator~\cite{JohnsonLippmann1950,KhachidzeKhelashvili2006}, which
commutes with the Dirac-Coulomb Hamiltonian and connects $\kappa$ with $-\kappa$;
at fixed $m_j$ the field therefore acts in manifolds of dimension one or two. The
\FV\ theory has the same levels, each carried twice with opposite $\tau_5$
norm~\eqref{eq:fvpm}, and the roots of its first-order problem are the Dirac roots
(\sect{sec:fvdoubled}). As far as the Stark effect is concerned, the three
theories therefore fall into two classes, nonrelativistic and relativistic, and the limits $Z\alpha\to0$ at fixed $E_0$ and $E_0\to0$ at fixed $Z\alpha$ do not commute: whenever a level has a fine-structure partner, the relativistic first-order roots at $Z\alpha\to0$ differ from the Schr\"odinger roots and the relativistic second-order shift diverges as $(Z\alpha)^{-2}$.
\Sectsr{sec:first}{sec:partner} make this statement quantitative.

\subsection{First order: closed forms and a second-moment sum rule}\label{sec:first}

\begin{figure}[tbp]
\centering
\includegraphics[width=\columnwidth]{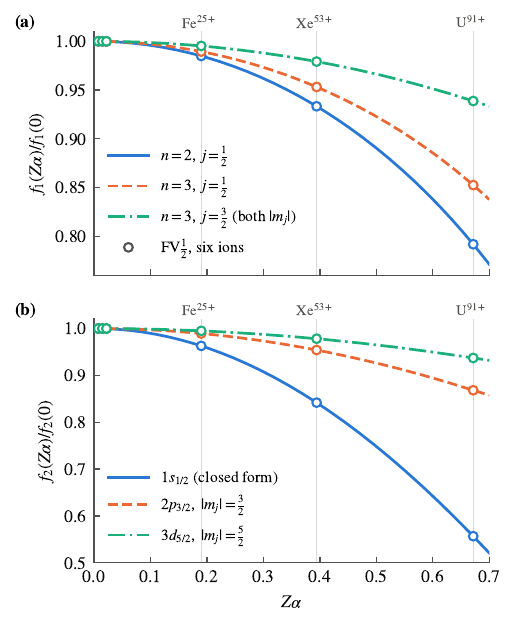}
\caption{Dependence on $Z\alpha$ of the relativistic shifts that remain finite as $Z\alpha\to0$. (a)~First-order roots, normalized to their values at $Z\alpha=0$ ($\sqrt3$, $3\sqrt2$ and
$9/(2\sqrt5)$ or $3/(2\sqrt5)$), from the closed forms~\eqref{eq:f1closed}. (b)~Second-order shifts of the levels without a fine-structure partner, normalized to the Schr\"odinger values $-9/4$, $-78$ and $-1377/2$; the $1s_{1/2}$ curve is the closed form of Szmytkowski~\cite{Szmytkowski1997}, and the others are Dalgarno--Lewis results. Lines: Dirac; open circles: \FV\ values of \tabl{tab:ions} for the six ions (H, He$^+$ and Li$^{2+}$ lie near $Z\alpha=0$).}
\label{fig:fig1}
\end{figure}

\paragraph{Closed forms.} The two members of a two-dimensional manifold share $\gamma$,
$N$ and the decay constant $Z/N$, so the radial integrals in \eq{eq:diracme} reduce to
ratios of $\Gamma$ functions whose arguments differ by integers, and the roots are
algebraic in $x\equiv Z\alpha$. With $\gamma_k\equiv(k^2-x^2)^{1/2}$,
\begin{widetext}
\begin{equation}\label{eq:f1closed}
\frac{ZE^{(1)}}{E_0}=\pm
\begin{cases}
\tfrac12(1+\gamma_1)\sqrt{1+2\gamma_1}, & n=2,\ j=\tfrac12,\\[2pt]
(2+\gamma_1)\sqrt{1+\gamma_1}, & n=3,\ j=\tfrac12,\\[2pt]
\tfrac{3}{10}(1+\gamma_2)\sqrt{1+2\gamma_2}, & n=3,\ j=\tfrac32,\ |m_j|=\tfrac32,\\[2pt]
\tfrac{1}{10}(1+\gamma_2)\sqrt{1+2\gamma_2}, & n=3,\ j=\tfrac32,\ |m_j|=\tfrac12 .
\end{cases}
\end{equation}
\end{widetext}
The Taylor expansions of \eq{eq:f1closed} are the series of \tabl{tab:table2rev}, their
values reproduce every first-order entry of \tabl{tab:ions}, and the \FV\ roots of
$2Y$ coincide with them to the working precision. The two $j=\tfrac32$ roots
are in the ratio $3 : 1$ at every $Z\alpha$, which is the ratio of the angular factors
$A_{\kappa'\kappa}$ of \eq{eq:diracme}, because both are built on the same radial
functions. \Fig{fig:fig1}(a) shows the three distinct functions normalized to
their values at $Z\alpha=0$.

Three features distinguish the relativistic from the nonrelativistic linear effect.
(i)~The relativistic roots are smaller by a finite factor that survives
$Z\alpha\to0$: $\sqrt3$ instead of $3$ at $n=2$, and $3\sqrt2=4.2426$ instead of $9$
for the extreme root at $n=3$ (in units of $E_0/Z$). (ii)~The pattern changes:
nonrelativistically the $n=3$ shell has the roots $\pm9$ and $\pm\tfrac92$ and a sixfold zero,
ordered by the parabolic quantum number $k$; relativistically it has the roots
$\pm3\sqrt2$, $\pm9/(2\sqrt5)$ and $\pm3/(2\sqrt5)$ and a sixfold zero, ordered by
$(j,|m_j|)$, and the $j=\tfrac52$ level is not shifted at first order. The two patterns are
not deformations of one another. (iii)~The remaining relativistic corrections (beyond this finite factor) vanish as $Z\alpha\to0$ and are small until $Z$ is large: from \eq{eq:f1closed} the three distinct roots are reduced by $2.2\times10^{-5}$, $1.6\times10^{-5}$ and
$7.1\times10^{-6}$ (relative) in hydrogen, and by $20.8\%$, $14.8\%$ and $6.1\%$ in U$^{91+}$.
The $j=\tfrac12$ manifolds, built on the penetrating $s_{1/2}$ and $p_{1/2}$
orbitals, are affected most.

\paragraph{Where the missing linear strength goes.} Feature~(i) does not mean that
part of the linear Stark effect is lost. We present the argument within the Dirac theory; since the \FV\ roots and shifts are the same, the conclusions hold for it as well. Let $\Pi$ project onto the states of a shell with a given $m_j$, and call $\mathrm{Tr}\,(\Pi\hat H_I\Pi)^2$, the sum of the squared moduli of all intra-shell matrix elements of $\hat H_I$, the first-order Stark strength of this sector. The eigenvalues of $\Pi(\hat H_0+\hat H_I)\Pi$ are
$E^{(0)}_i+E^{(1)}_i+E^{(2)}_{i,\rm intra}+\Or(E_0^3)$, where $E^{(1)}_i\propto E_0$ are the first-order roots and $E^{(2)}_{i,\rm intra}\propto E_0^2$ is the part of \eq{eq:E2deg} due to the other levels of the same shell (\sect{sec:partner}). The first two moments of these eigenvalues are fixed by the matrix: because the diagonal elements of $\hat H_I$ vanish by parity, $\mathrm{Tr}\,\Pi(\hat H_0+\hat H_I)\Pi=\mathrm{Tr}\,\Pi\hat H_0\Pi$ and $\mathrm{Tr}\,[\Pi(\hat H_0+\hat H_I)\Pi]^2=\mathrm{Tr}\,(\Pi\hat H_0\Pi)^2+\mathrm{Tr}\,(\Pi\hat H_I\Pi)^2$
exactly. The terms of second order in $E_0$ give, at every $Z\alpha$,
\begin{gather}
\sum_iE^{(2)}_{i,\rm intra}=0,\label{eq:moment1}\\
\sum_i\Bigl[\bigl(E^{(1)}_i\bigr)^2+2E^{(0)}_iE^{(2)}_{i,\rm intra}\Bigr]=\mathrm{Tr}\,(\Pi\hat H_I\Pi)^2.\label{eq:moment2}
\end{gather}
The $(Z\alpha)^{-2}$ term of \eq{eq:moment1} is the sum rule~\eqref{eq:sumrule1}. In \eq{eq:moment2},
relation~\eqref{eq:moment1} removes the rest energy and the Bohr energy, which are common to the shell,
so that only the fine-structure energies remain, and as $Z\alpha\to0$ these multiply the singular terms
$c_{-2}/(Z\alpha)^2$; the right-hand side becomes the Stark strength of the nonrelativistic shell,
$\sum_i(E^{(1)}_{i,\rm S})^2$. In the units~\eqref{eq:universal},
\begin{equation}\label{eq:secondmoment}
\sum_{i\in(n,m_j)}\Bigl[f_{1,i}(0)^2+2\,\phi_{n j_i}\,c_{-2,i}\Bigr]
=\sum_{i\in(n,m_j)}\bigl(f_{1,i}^{\rm S}\bigr)^2,
\end{equation}
where $\phi_{nj}=-\bigl[n/(j+\frac12)-\frac34\bigr]/(2n^4)$, so that $\phi_{nj}(Z\alpha)^2$ is the fine-structure term of $(E_{n,j}-c^2)/Z^2$ from \eq{eq:diracE},
the sums run over all states of the shell with the given $m_j$, and $f^{\rm S}_1$ are the
Schr\"odinger roots of \tabl{tab:schr}. Because $\sum_ic_{-2,i}=0$, only
the fine-structure intervals enter. With the coefficients of \tabl{tab:table2rev} both sides
equal $18$ for $n=2$, $m_j=\tfrac12$; $405/2$ for $n=3$, $m_j=\tfrac12$; and $81/2$ for
$n=3$, $m_j=\tfrac32$. The sum rules~\eqref{eq:sumrule1} and~\eqref{eq:secondmoment} are thus the
$Z\alpha\to0$ forms of the first two spectral moments of the shell: the first-order Stark strength
of the nonrelativistic shell is conserved, carried partly by the relativistic first-order
roots and partly by the couplings between the fine-structure levels, which reappear at second order, enhanced by $(Z\alpha)^{-2}$. At $n=2$ the sector sum $18$ is twice $3^2=(\sqrt3)^2+(\sqrt6)^2$, the quadrature sum of the matrix elements~\eqref{eq:schr2me}.

\paragraph{Relation to earlier work.} Since $2s_{1/2}$ and $2p_{1/2}$ are the only members of their manifold, the $n=2$ root
is $Z|v_1|$, the modulus of the coupling of \sect{sec:fvlevelshift}. Hillery and Mohr~\cite{HilleryMohr1980} evaluated $v_1$
and $v_3$ numerically, as $Z|v_1|=\sqrt3[1-0.4167x^2F_5]$ and $Z|v_3|=\sqrt6[1-0.1667x^2F_6]$, with $F_5$ and $F_6$ given
graphically for $Z\le60$. \Eq{eq:f1closed} gives $F_5$ in closed form,
\begin{equation}\label{eq:F5}
F_5=\frac{12}{5x^2}\Bigl[1-\frac{1+\gamma_1}{2}\Bigl(\frac{1+2\gamma_1}{3}\Bigr)^{1/2}\Bigr]=1+\frac{11}{60}x^2+\Or(x^4),
\end{equation}
with $F_5=1.0032$, $1.0163$ and $1.0389$ at $Z=18$, $40$ and $60$. Similarly, the exact $v_3$,
$Z|v_3|=\sqrt6\,[1-\frac16x^2-(\frac{9}{512}+\frac{\pi^2}{192})x^4+\Or(x^6)]$, i.e., $F_6=1+(\frac{27}{256}+\frac{\pi^2}{32})x^2+\dots$,
gives $F_6=1.0072$, $1.0372$ and $1.0899$. Both agree with the curves of Ref.~\cite{HilleryMohr1980} to within reading accuracy. With $\alpha=1/137$, as throughout,
the line strengths $6(Zv_1)^2$ and $6(Zv_3)^2$ agree with the relativistic values of Younger and Weiss~\cite{YoungerWeiss1975}
for $Z\le90$ to within $0.04\%$. To our knowledge, \eq{eq:f1closed} has not been stated before, and the quench analyses use the numerical values of \citeref{HilleryMohr1980}; matrix elements between states of the same $n$ are, however, elementary combinations of $\Gamma$ functions, accessible also through general relations for relativistic radial integrals~\cite{Shabaev1991}.

\subsection{Second order: levels without a fine-structure partner}\label{sec:regular}

For $n=1$ and for $(n,j,|m_j|)=(2,\tfrac32,\tfrac32)$ and $(3,\tfrac52,\tfrac52)$, the
relativistic second-order shift is a smooth deformation of the nonrelativistic one. Including the $x^4$ terms of \tabl{tab:table2rev},
\begin{widetext}
\begin{equation}\label{eq:regular}
\frac{Z^4E^{(2)}}{E_0^2}=
\begin{cases}
-\frac94+\frac73x^2-\bigl(\frac{31}{192}+\frac{\pi^2}{96}\bigr)x^4+\Or(x^6), & 1s_{1/2},\\[3pt]
-78+\frac{93}{4}x^2-\bigl(\frac{85}{576}+\frac{\pi^2}{12}\bigr)x^4+\Or(x^6), & 2p_{3/2},\ |m_j|=\tfrac32,\\[3pt]
-\frac{1377}{2}+\frac{1359}{14}x^2+\bigl(\frac{32549}{44800}-\frac{9\pi^2}{32}\bigr)x^4+\Or(x^6), & 3d_{5/2},\ |m_j|=\tfrac52,
\end{cases}
\end{equation}
\end{widetext}
where the $x^2$ and $x^4$ coefficients were identified from the numerical results as described in the
caption of \tabl{tab:table2rev}. The leading terms are the Schr\"odinger
values for the parabolic states $(k,m)=(0,0)$, $(0,\pm1)$ and $(0,\pm2)$. For the
ground state the full function is known in closed form from the Sturmian expansion
of the Dirac--Coulomb Green function~\cite{Szmytkowski1997}: our Dalgarno--Lewis
values agree with it to 45 significant figures for all six values in $Z$ of
\tabl{tab:ions}, and its expansion, $\alpha_d=\frac{9}{2Z^4}\bigl[1-\frac{28}{27}x^2+\bigl(\frac{31}{432}+\frac{\pi^2}{216}\bigr)x^4+\dots\bigr]$, confirms the $x^4$ coefficient of \eq{eq:regular} independently.

The relativistic correction is positive in all three cases: the relativistic contraction
makes the atom stiffer. Its relative size falls quickly along the sequence of nodeless
states, $c_2/c_0=-28/27$, $-31/104$ and $-151/1071$, and at U$^{91+}$ the scaled
shift is reduced to $0.557$, $0.868$ and $0.937$ times its nonrelativistic value
(\fig{fig:fig1}(b)); the $2p_{3/2}$ and $3d_{5/2}$ orbitals are kept away from
the nucleus by the centrifugal barrier and feel the contraction far less than the $1s_{1/2}$ orbital does.

\subsection{Second order: levels with a fine-structure partner}\label{sec:partner}

\paragraph{Decomposition.} For the other seven $(n,j,|m_j|)$ levels, the second-order shift
contains the singular terms $c_{-2}/x^2$ of \tabl{tab:table2rev}. We separate the contribution of the other levels $(n,j')$ of the same shell (states $\beta$ with the same $m_j$) from that of all other states, $f_2^{\rm extra}$:
\begin{gather}
f_2=f_2^{\rm intra}+f_2^{\rm extra},\label{eq:decomp}\\
f_2^{\rm intra}\bigl(n,j,|m_j|\bigr)=\frac{Z^4}{E_0^2}\sum_{j'\neq j,\,\beta}
\frac{|\langle\psi_{\beta}|\hat H_I|\bar\psi^{(0)}\rangle|^2}{E_{n,j}-E_{n,j'}},\label{eq:intradef}
\end{gather}
Both matrix elements and denominators in $f_2^{\rm intra}$ are known in closed form from \eqsss{eq:diracE}{eq:diracfg}{eq:diracme}, so $f_2^{\rm intra}$ is an explicit function of $Z\alpha$ built from $\Gamma$ functions; $f_2^{\rm extra}$ is the Galerkin result minus $f_2^{\rm intra}$. The decomposition has three properties:

(a)~$f_2^{\rm intra}$ contains all the singular terms. For $n=2$, where the
$2s_{1/2}$--$2p_{3/2}$ element $v_3E_0$ is the only intra-shell coupling,
$f_2^{\rm intra}(2,\tfrac12,\tfrac12)=-\tfrac12(Zv_3)^2Z^2/\Delta_2$, with $\Delta_2$ the interval~\eqref{eq:Delta2}, and
\begin{equation}\label{eq:intra2}
f_2^{\rm intra}\bigl(2,\tfrac12,\tfrac12\bigr)=-\frac{96}{x^2}+92+\Bigl(\pi^2-\frac{409}{24}\Bigr)x^2+\Or(x^4),
\end{equation}
and $f_2^{\rm intra}(2,\tfrac32,\tfrac12)=-2f_2^{\rm intra}(2,\tfrac12,\tfrac12)$,
the leading terms being those of \eq{eq:singular52} below. The coefficients of $f_2^{\rm intra}$ for
all seven levels are listed in \tabl{tab:decomp}.

(b)~$f_2^{\rm extra}$ is regular at $Z\alpha=0$, and its limit is a nonrelativistic
quantity: the expectation value of the second-order operator
$\hat M^{(n)}=\sum_{k\notin\{n\}}\hat H_I|k\rangle\langle k|\hat H_I/(E_n-E_k)$
of \eq{eq:Mmatrix}, with the whole shell excluded, in the $Z\alpha\to0$ limit of the
relativistic correct zeroth-order state. The operator $\hat M^{(n)}$ is spin independent and even,
so it is diagonal in $m$ and couples $l$ only to $l$ and $l\pm2$. The Dalgarno--Lewis
equation~\eqref{eq:DL} with the shell projected out gives it exactly; in units of
$E_0^2/Z^4$ its elements $M_{nl,m}$ are
$M_{2s,0}=-60$, $M_{2p,0}=-108$, $M_{2p,\pm1}=-78$ for $n=2$, and
$M_{3s,0}=-\frac{2025}{4}$, $M_{3p,0}=-810$, $M_{3d,0}=-\frac{2349}{2}$,
$M_{3s3d,0}=-\frac{243\sqrt2}{4}$, $M_{3p,\pm1}=-567$, $M_{3d,\pm1}=-1053$,
$M_{3d,\pm2}=-\frac{1377}{2}$ for $n=3$. In the parabolic
basis their diagonal elements are the entries of \tabl{tab:schr}; for example
$\frac12(M_{2s,0}+M_{2p,0})=-84$. For a two-dimensional relativistic manifold the correct
zeroth-order states at $Z\alpha\to0$ are equal-weight combinations of $\kappa$ and $-\kappa$, whose cross terms in
$\hat M^{(n)}$ vanish by parity, so that
$f_2^{\rm extra}(0)=\frac1\nu\sum_{\kappa}\sum_{m_s}\bigl[C(l\tfrac12j;m\,m_s\,m_j)\bigr]^2M_{nl,m}$,
with $\nu=1$ or $2$ the dimension of the manifold, $l=l(\kappa)$ and $m=m_j-m_s$.
For $(2,\tfrac12,\tfrac12)$ this is $\frac12\bigl[M_{2s,0}+\frac13M_{2p,0}+\frac23M_{2p,\pm1}\bigr]=-74$,
and for $(2,\tfrac32,\tfrac12)$ it is $\frac23M_{2p,0}+\frac13M_{2p,\pm1}=-98$. Adding the
constant terms of $f_2^{\rm intra}$ reproduces the coefficients $c_0$ of \tabl{tab:table2rev},
$92-74=18$ and $-184-98=-282$, and \tabl{tab:decomp} shows that the same holds
exactly for all seven levels. Every $c_0$ of of these seven levels is thereby derived
analytically, independently of the Galerkin solution. Each such $c_0$ is the sum of two parts: the nonrelativistic second-order energy $\langle\hat M^{(n)}\rangle$ of the relativistic
$(j,m_j)$ state, and the constant term $c_0^{\rm intra}$ of the intra-shell part, which arises from the $\Or\bigl((Z\alpha)^2\bigr)$ corrections to both the intra-shell coupling and the fine-structure interval.

(c)~The terms in $\pi^2$ are intra-shell, and their coefficients follow from $c_{-2}$. The $x^2$
coefficients $c_2$ of the seven levels have the form $a+b\pi^2$ (\tabl{tab:table2rev}). Consider
the coupling of a level with $|\kappa|=k$ to its partner with $|\kappa|=k+1$ in the same shell. In the
radial integral and in the normalization constants the $\Gamma$ functions combine into
$\Gamma(u)^2/[\Gamma(w)\Gamma(w')]$, with $u=\gamma+\gamma'+q$ ($q$ an integer), $w=2\gamma+1$ and
$w'=2\gamma'+1$, multiplied by factors that are rational in $\gamma$ and $\gamma'$ and by powers of the decay
constants; since all states of the shell share the decay constant $Z/n$ at $Z\alpha=0$, the latter reduce
to integer powers of $Z/n$ times power series in $x^2$ with rational coefficients, and no logarithm of
$Z\alpha$ appears. The arguments $u$, $w$, $w'$ are integers plus increments of order $x^2$, related by
$\delta_u=(\delta_w+\delta_{w'})/2$. When $\ln\Gamma$ is expanded to second order, the terms containing Euler's constant $\gamma_{\rm E}$, which come from the digamma function $\psi=\Gamma'/\Gamma$, are proportional to $2\delta_u-\delta_w-\delta_{w'}$ and cancel, and the trigamma function at positive integer argument, $\psi'(\ell)=\pi^2/6-\sum_{t=1}^{\ell-1}t^{-2}$~\cite{DLMF}, contributes the $\pi^2$-dependent part $\frac{\pi^2}{12}(2\delta_u^2-\delta_w^2-\delta_{w'}^2)=-\frac{\pi^2}{24}(\delta_w-\delta_{w'})^2=-\pi^2x^4/[24k^2(k+1)^2]$, because $\gamma=k-x^2/(2k)+\Or(x^4)$ and $\gamma'=k+1-x^2/[2(k+1)]+\Or(x^4)$. The squared coupling therefore carries the factor $1-\pi^2x^4/[24k^2(k+1)^2]$ times a series with rational coefficients, the fine-structure interval carries no $\pi^2$, and
\begin{equation}\label{eq:bformula}
b=-\sum_{\rm partners}\frac{c_{-2}^{(\rm p)}}{24k^2(k+1)^2},
\end{equation}
where $c_{-2}^{(\rm p)}$ is the contribution of the partner to $c_{-2}$ and $k$ is the smaller of the two values of $|\kappa|$. For $n=2$ ($k=1$) this gives $b=96/96=1$ and $-192/96=-2$. For $(3,\frac32,\frac12)$ the partners $j=\frac12$ ($k=1$) and $j=\frac52$ ($k=2$) contribute $+3159$ and $-19683/5$ to $c_{-2}=-3888/5$, and $b=-3159/96+19683/(5\cdot864)=-567/20$. \Eq{eq:bformula} gives $b$ for all seven levels of \tabl{tab:table2rev}, and the remainders $c_2^{\rm extra}=c_2-c_2^{\rm intra}$ are rational to the 26 significant figures of the identification (\tabl{tab:decomp}), so that the $\pi^2$ terms of $c_2$ are
entirely intra-shell. We note that $\pi^2$ also appears in the $x^4$ coefficients
of \eq{eq:regular}, which are of the same relative order $(Z\alpha)^4$; there the intermediate states belong to other shells, the argument above does not apply, and we have not traced the origin of these terms.

\begin{table*}[tbp]
\caption{Decomposition~\eqref{eq:decomp} of the second-order coefficients of the levels with a fine-structure partner, $Z^4E^{(2)}/E_0^2=c_{-2}/x^2+c_0+c_2x^2+\Or(x^4)$, $x=Z\alpha$. The singular coefficient $c_{-2}$ is purely intra-shell; $c_0=c_0^{\rm intra}+\langle\hat M^{(n)}\rangle$, where $\langle\hat M^{(n)}\rangle=f_2^{\rm extra}(0)$ is the expectation value of the nonrelativistic second-order operator of the shell in the relativistic correct zeroth-order state; and $c_2=c_2^{\rm intra}+c_2^{\rm extra}$. The intra-shell coefficients follow from the closed form of $f_2^{\rm intra}$, the values of $\langle\hat M^{(n)}\rangle$ from the elements of $\hat M^{(n)}$ given in the text, and the coefficients of $\pi^2$ in $c_2^{\rm intra}$ follow from $c_{-2}$ through \eq{eq:bformula}; the sums $c_0$ and $c_2$ are the coefficients of \tabl{tab:table2rev}.}
\label{tab:decomp}
\begin{ruledtabular}
\begin{tabular}{ccrrrrll}
$n$ & $(j,|m_j|)$ & $c_{-2}$ & $c_0^{\rm intra}$ & $\langle\hat M^{(n)}\rangle$ & $c_0$ & $c_2^{\rm intra}$ & $c_2^{\rm extra}$ \\
\hline
2 & $(\frac12,\frac12)$ & $-96$ & $92$ & $-74$ & $18$ & $\pi^2-\frac{409}{24}$ & $\frac{847}{8}$ \\[3pt]
2 & $(\frac32,\frac12)$ & $192$ & $-184$ & $-98$ & $-282$ & $\frac{409}{12}-2\pi^2$ & $\frac{89}{6}$ \\[3pt]
3 & $(\frac12,\frac12)$ & $-3159$ & $\frac{43659}{16}$ & $-\frac{4617}{8}$ & $\frac{34425}{16}$ & $\frac{1053}{32}\pi^2-\frac{14979}{40}$ & $\frac{7653}{8}$ \\[3pt]
3 & $(\frac32,\frac32)$ & $-\frac{13122}{5}$ & $\frac{143613}{200}$ & $-\frac{3321}{5}$ & $\frac{10773}{200}$ & $\frac{243}{80}\pi^2-\frac{14391}{250}$ & $\frac{113337}{400}$ \\[3pt]
3 & $(\frac32,\frac12)$ & $-\frac{3888}{5}$ & $-\frac{165159}{100}$ & $-\frac{9153}{10}$ & $-\frac{256689}{100}$ & $\frac{288129}{1000}-\frac{567}{20}\pi^2$ & $\frac{81993}{400}$ \\[3pt]
3 & $(\frac52,\frac32)$ & $\frac{26244}{5}$ & $-\frac{143613}{100}$ & $-\frac{9801}{10}$ & $-\frac{241623}{100}$ & $\frac{14391}{125}-\frac{243}{40}\pi^2$ & $\frac{18747}{175}$ \\[3pt]
3 & $(\frac52,\frac12)$ & $\frac{39366}{5}$ & $-\frac{430839}{200}$ & $-\frac{11259}{10}$ & $-\frac{656019}{200}$ & $\frac{43173}{250}-\frac{729}{80}\pi^2$ & $\frac{78507}{700}$ \\[3pt]
\end{tabular}
\end{ruledtabular}
\end{table*}

\paragraph{Sum rules.} By \eq{eq:moment1} the sum of $f_2^{\rm intra}$ over the states of a given
$(n,m_j)$ vanishes identically in $Z\alpha$: two levels of a shell shift each other by equal and
opposite amounts. Three consequences follow. The singular coefficients cancel, which is
\eq{eq:sumrule1}; the $\pi^2$ parts of $c_2$ cancel as well, $2(1)-2=0$ for $n=2$ and
$2\cdot\frac{1053}{32}-2\cdot\frac{567}{20}-\frac{729}{80}=0$, $2\cdot\frac{243}{80}-\frac{243}{40}=0$
for $n=3$, as \eq{eq:bformula} requires; and the sum of $f_2$ over each $(n,m_j)$ equals that of the
regular function $f_2^{\rm extra}$, whose value at $Z\alpha=0$ is the trace of $\hat M^{(n)}$ over the same
states, i.e., the corresponding nonrelativistic sum. This sharpens \eq{eq:sumrule2} into a relation that holds sector by sector: for $n=2$, $m_j=\frac12$, $2(18)-282=-246=2(-84)-78$, and for $n=3$, $m_j=\frac12$ and $\frac32$, both sides equal $-16443/4$ and $-4617/2$, respectively.

\begin{figure}[tbp]
\centering
\includegraphics[width=\columnwidth]{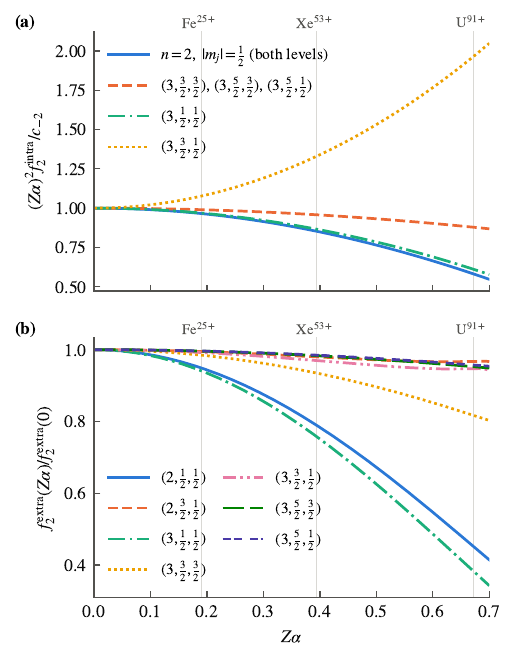}
\caption{The two parts~\eqref{eq:decomp} of the second-order shift of the levels with a fine-structure partner. (a)~Intra-shell part, multiplied by $(Z\alpha)^2$ and normalized to its singular coefficient $c_{-2}$. Levels whose only intra-shell coupling is the same radial matrix element give the same curve: the two levels of $n=2$, $|m_j|=\frac12$ ($2s_{1/2}$--$2p_{3/2}$ coupling); and $(3,\frac32,\frac32)$, $(3,\frac52,\frac32)$ and $(3,\frac52,\frac12)$ ($3p_{3/2}$--$3d_{5/2}$ coupling). For $(3,\frac32,\frac12)$ the couplings to the levels below ($+3159$) and above ($-19683/5$) partly cancel in $c_{-2}$, and the normalized curve rises. (b)~Regular part, normalized to its value $\langle\hat M^{(n)}\rangle$ at $Z\alpha=0$ (\tabl{tab:decomp}). The Dirac and \FV\ theories give identical curves.}
\label{fig:fig2}
\end{figure}

\paragraph{Size of the effect.} In hydrogen the scaled shifts of the partner levels reach
$10^6$--$10^8$ (\tabl{tab:ions}), compared to $10^2$--$10^3$ nonrelativistically, because the energy
denominator is the fine-structure interval, of order $Z^4\alpha^2$, rather than a shell spacing
of order $Z^2$. This is not a breakdown of the relativistic theories but a statement about the
field at which the second order competes with the first (\sect{sec:validity}). \Fig{fig:fig2}
shows the two parts separately: the singular part normalized to its leading term, and the regular
part normalized to its value at $Z\alpha=0$. The regular part is reduced by the relativistic contraction, like the
shifts of \sect{sec:regular}, and again most strongly for the manifolds built on $s_{1/2}$ and $p_{1/2}$ orbitals:
at U$^{91+}$ it retains $0.45$ and $0.38$ of its nonrelativistic value for $(2,\frac12,\frac12)$ and $(3,\frac12,\frac12)$,
and between $0.82$ and $0.97$ for the other five levels.

\paragraph{The continuum.} The second-order shift is dominated, but not exhausted, by the bound
intermediate states: in the nonrelativistic theory a sum restricted to the discrete spectrum recovers $81.4\%$ of the ground-state result, $94.8\%$ and $96.4\%$ at $n=2$ and $97.6$--$99.1\%$ at $n=3$
(\tabl{tab:schr}). In the relativistic theories the spectral sum runs in addition over the
negative-energy states; the Dalgarno--Lewis solution includes them, together with the positive-energy
continuum, and their contribution is part of $f_2^{\rm extra}$. Since the \FV\ and Dirac shifts are compared
to tens of significant figures, and the sum rules above test the coefficients exactly, a truncated
spectral sum would have been useless for this purpose.

\subsection{The weak-field/strong-field crossover}\label{sec:crossover}

The nonrelativistic roots $\pm3E_0/Z$ and the relativistic roots $\pm\sqrt3E_0/Z$ are the two limits of a single crossover, governed by the ratio of the Stark coupling to the fine-structure interval $\Delta_2$. We retain the $n=2$, $m_j=\frac12$ sector $\{2s_{1/2},2p_{1/2},2p_{3/2}\}$ and diagonalize, without expanding in $E_0$, the matrix
\begin{equation}\label{eq:threelevel}
H=\begin{pmatrix}0&v_1E_0&v_3E_0\\ v_1E_0&0&0\\ v_3E_0&0&\Delta_2\end{pmatrix},
\end{equation}
with $Zv_1\to-\sqrt3$ and $Zv_3\to-\sqrt6$ as $Z\alpha\to0$, where $v_1$ and $v_3$ are the exact Dirac matrix elements~\eqref{eq:diracme} in the phase convention
of \eq{eq:diracfg} ($Zv_1$ is the $n=2$ root of \eq{eq:f1closed}; the $2p_{1/2}$--$2p_{3/2}$ element
vanishes by parity). The roots are $\{+1,0,-1\}Z(v_1^2+v_3^2)^{1/2}E_0/Z$ for $\Delta_2\to0$, which tend to
$\{+3,0,-3\}E_0/Z$ as $Z\alpha\to0$, and $\pm\sqrt3E_0/Z$ for
$E_0\to0$. Expanding the latter to second order in $E_0$ reproduces the singular coefficients
of \eq{eq:E2n2},
\begin{equation}\label{eq:singular52}
-\frac{3E_0^2}{Z^2\Delta_2}=-\frac{96}{Z^4(Z\alpha)^2}E_0^2,\qquad
+\frac{6E_0^2}{Z^2\Delta_2}=+\frac{192}{Z^4(Z\alpha)^2}E_0^2,
\end{equation}
so these terms are the leading edge of the return to the nonrelativistic regime; the sum
rule~\eqref{eq:sumrule1} is the conservation of the trace of \eq{eq:threelevel}, and \eq{eq:secondmoment}
the conservation of the trace of its square. The crossover is controlled by
$|v_3|E_0/\Delta_2\simeq\sqrt6E_0/(Z\Delta_2)$.

\begin{table*}[tbp]
\caption{Lowest root of the three-level problem~\eqref{eq:threelevel} in units of $E_0/Z$, as the field is taken across the
fine-structure interval $\Delta_2$ (exact Dirac matrix elements and exact $\Delta_2$; 1~a.u.\ of field $=5.142\times10^9$~V\,cm$^{-1}$).
The column headed $E_0\to0$ gives the perturbative result $-\frac12(1+\gamma_1)\sqrt{1+2\gamma_1}=Zv_1$ of \eq{eq:f1closed}; at large fields
the root approaches the strong-field limit $-Z(v_1^2+v_3^2)^{1/2}$ of \eq{eq:threelevel}, which tends to the nonrelativistic
$-3$ as $Z\alpha\to0$ ($-2.973$, $-2.878$ and $-2.607$ for Fe$^{25+}$, Xe$^{53+}$ and U$^{91+}$). The field $E_{\rm mid}$ is the one at which
the root lies midway between these two limits. The entries at $100\Delta_2$ for Fe$^{25+}$, Xe$^{53+}$ and U$^{91+}$
($E_0/Z^3\approx0.004$--$0.02$), and $E_{\rm mid}$ for U$^{91+}$, lie beyond the over-barrier field $\sim Z^3/256$ (\sect{sec:validity}),
where the couplings to other shells, neglected in \eq{eq:threelevel}, are no longer small, and these entries only illustrate the model. The notation $a(b)$ means $a\times10^b$.}
\label{tab:crossover}
\begin{ruledtabular}
\begin{tabular}{lllrrrrrrr}
ion & $\Delta_2$ (a.u.) & $\Delta_2$ (V\,cm$^{-1}$) & $E_0\to0$ & $0.01\Delta_2$ & $0.1\Delta_2$ & $\Delta_2$ & $10\Delta_2$ & $100\Delta_2$ & $E_{\rm mid}/\Delta_2$ \\
\hline
H & 1.665($-6$) & 8.56(3) & $-1.7320$ & $-1.7617$ & $-2.0000$ & $-2.7177$ & $-2.9672$ & $-2.9966$ & 0.323 \\
He$^{+}$ & 2.664($-5$) & 1.370(5) & $-1.7319$ & $-1.7468$ & $-1.8745$ & $-2.5186$ & $-2.9354$ & $-2.9932$ & 0.646 \\
Li$^{2+}$ & 1.349($-4$) & 6.937(5) & $-1.7317$ & $-1.7417$ & $-1.8285$ & $-2.3772$ & $-2.9045$ & $-2.9897$ & 0.968 \\
Fe$^{25+}$ & 7.784($-1$) & 4.003(9) & $-1.7059$ & $-1.7070$ & $-1.7172$ & $-1.8157$ & $-2.3988$ & $-2.8893$ & 8.46 \\
Xe$^{53+}$ & 1.569(1) & 8.069(10) & $-1.6165$ & $-1.6170$ & $-1.6217$ & $-1.6682$ & $-2.0419$ & $-2.7097$ & 18.1 \\
U$^{91+}$ & 1.679(2) & 8.634(11) & $-1.3714$ & $-1.3717$ & $-1.3741$ & $-1.3980$ & $-1.6174$ & $-2.3246$ & 33.8 \\
\end{tabular}
\end{ruledtabular}
\end{table*}

\Tabl{tab:crossover} follows the lowest root across the crossover. Measured between its two limits,
$Zv_1$ and $-Z(v_1^2+v_3^2)^{1/2}$, the root reaches the midpoint at $E_{\rm mid}=0.323\,Z\Delta_2$ for
H, He$^+$ and Li$^{2+}$ and at $0.326$, $0.335$ and $0.368\,Z\Delta_2$ for Fe$^{25+}$, Xe$^{53+}$ and U$^{91+}$:
in the variable $E_0/(Z\Delta_2)$ the crossover is nearly universal, as the scaling of \eq{eq:threelevel}
implies for $Z\alpha\ll1$. For hydrogen the scale $\Delta_2$ corresponds to $8.6$~kV\,cm$^{-1}$ and the midpoint to
$2.8$~kV\,cm$^{-1}$, both ordinary laboratory fields: within the Dirac (and \FV) model the relativistic value
$\sqrt3$ applies below about $0.1$~kV\,cm$^{-1}$, the nonrelativistic value $3$ above about
$10^3$~kV\,cm$^{-1}$, and in between the three-level problem must be diagonalized. In real hydrogen the
$2s_{1/2}$--$2p_{1/2}$ degeneracy is lifted by the Lamb shift~\cite{Lamb1947} ($\approx1058$~MHz, about one-tenth of $\Delta_2$), which makes the response of these two states quadratic at the lowest fields; the $\sqrt3$
regime is a property of the Dirac and \FV\ models, not an effect we propose to measure. When the radiative level energies (complex because of the level widths) are placed on its diagonal, \eq{eq:threelevel} becomes the model on which the analysis of the field-quenched decay of the $2s$ level rests (\exteq{20} of \citeref{HilleryMohr1980}); in the \FV\ theory the shifts must be
introduced as in \sect{sec:fvlevelshift}. For U$^{91+}$ the midpoint, $5.7\times10^3$~a.u.\ or about $3\times10^{13}$~V\,cm$^{-1}$, lies beyond the
over-barrier field estimate of \sect{sec:validity} ($\approx3\times10^3$~a.u.), so the crossover is only formal there: in
any field in which the bound levels survive, the fine-structure-resolved result holds. The heavier the ion, the
more securely the relativistic linear Stark effect applies---the opposite of what the growth of the relativistic
corrections alone would suggest.

\subsection{Range of validity of the perturbation expansion}\label{sec:validity}

Remark~(iii) of \sect{sec:units} requires $|E^{(2)}|\ll|E^{(1)}|$ for the levels with a linear shift, i.e.,
$E_0\ll E_0^{\max}\equiv E_0|E^{(1)}/E^{(2)}|$. With \eq{eq:universal},
\begin{equation}\label{eq:E0max}
E_0^{\max}=\left|\frac{f_1(Z\alpha)}{f_2(Z\alpha)}\right|Z^3\simeq\frac{\sqrt3}{96}\,\alpha^2Z^5
\qquad\bigl(n=2,\ j=\tfrac12\bigr),
\end{equation}
the last form following from $f_1\to\sqrt3$ and $f_2\to-96/(Z\alpha)^2$. For the three levels that have a
fine-structure partner but no linear shift, $(2,\frac32,\frac12)$, $(3,\frac52,\frac32)$ and $(3,\frac52,\frac12)$,
the second-order shift must remain small compared with the interval $\Delta$ to the partner,
$|E^{(2)}|\ll\Delta$. \Tabl{tab:validity} evaluates both limits. In all seven cases the limit is
set by the fine-structure interval and grows as $\alpha^2Z^5$ for $Z\alpha\ll1$ (last column): it is the light ions,
not the heavy ones, whose perturbative window is narrow, because the fine structure appears in the energy
denominator. For hydrogen at $n=2$ the limit is $9.6\times10^{-7}$~a.u.\ $\simeq4.9$~kV\,cm$^{-1}$, of the same order
as the crossover field of \sect{sec:crossover}, as it must be; for U$^{91+}$ it is $6.8\times10^{3}$~a.u.\
$\simeq3.5\times10^{13}$~V\,cm$^{-1}$. The fine structure is not the only constraint. The simple over-barrier estimate of the field
beyond which field ionization~\cite{Maltsev2023} becomes rapid, $\sim Z^3/(16n^4)$~a.u., grows as $Z^3$; it lies far above the entries of \tabl{tab:validity} for the light ions, but for Xe$^{53+}$ and U$^{91+}$
($6\times10^2$ and $3\times10^3$~a.u.\ at $n=2$) it is comparable with them, and for the heaviest ions field
ionization, not the fine structure, bounds the perturbative regime. For the three levels without a fine-structure partner, field ionization and the mixing of neighboring shells are the only limit.

\begin{table*}[tbp]
\caption{Limits of the perturbative regime, in atomic units of field ($\alpha=1/137$; 1~a.u.\ $=5.142\times10^9$~V\,cm$^{-1}$).
Upper block: levels with a linear shift, $E_0^{\max}=E_0|E^{(1)}/E^{(2)}|$. Lower block: levels with a fine-structure partner but no
linear shift; the entries are the field at which $|E^{(2)}|$ equals the interval $\Delta$ to the partner. The last column is the leading form for
$Z\alpha\ll1$. For the levels without a fine-structure partner the expansion is limited by the mixing of neighboring shells
and by field ionization; the simple over-barrier estimate $\sim Z^3/(16n^4)$~a.u.\ (about $4\times10^6$~V\,cm$^{-1}$ for hydrogen $n=3$)
lies far above the entries for light ions but is comparable with them for Xe$^{53+}$ and U$^{91+}$. The notation $a(b)$ means $a\times10^b$.}
\label{tab:validity}
\begin{ruledtabular}
\begin{tabular}{cclllllll}
$n$ & $(j,|m_j|)$ & H & He$^{+}$ & Li$^{2+}$ & Fe$^{25+}$ & Xe$^{53+}$ & U$^{91+}$ & $Z\alpha\ll1$ \\
\hline
2 & $(\frac12,\frac12)$ & 9.6($-7$) & 3.1($-5$) & 2.3($-4$) & 1.1(1) & 4.3(2) & 6.8(3) & $\alpha^2Z^5/55.4$ \\
3 & $(\frac12,\frac12)$ & 7.2($-8$) & 2.3($-6$) & 1.7($-5$) & 8.6($-1$) & 3.5(1) & 6.2(2) & $\alpha^2Z^5/745$ \\
3 & $(\frac32,\frac32)$ & 4.1($-8$) & 1.3($-6$) & 9.9($-6$) & 4.8($-1$) & 1.8(1) & 2.6(2) & $\alpha^2Z^5/1304$ \\
3 & $(\frac32,\frac12)$ & 4.6($-8$) & 1.5($-6$) & 1.1($-5$) & 4.9($-1$) & 1.4(1) & 1.2(2) & $\alpha^2Z^5/1159$ \\
\hline
2 & $(\frac32,\frac12)$ & 6.8($-7$) & 2.2($-5$) & 1.7($-4$) & 8.4 & 3.7(2) & 8.9(3) & $\alpha^2Z^5/78.4$ \\
3 & $(\frac52,\frac32)$ & 4.1($-8$) & 1.3($-6$) & 9.9($-6$) & 4.9($-1$) & 2.0(1) & 3.1(2) & $\alpha^2Z^5/1304$ \\
3 & $(\frac52,\frac12)$ & 3.3($-8$) & 1.1($-6$) & 8.1($-6$) & 4.0($-1$) & 1.6(1) & 2.5(2) & $\alpha^2Z^5/1597$ \\
\end{tabular}
\end{ruledtabular}
\end{table*}

\subsection{Dirac versus \FV: what the agreement establishes}\label{sec:DvsFV}

\paragraph{Precision of the agreement.} The two relativistic theories agree level by level and order by
order (\sect{sec:fvresults}). At second order the two calculations follow separate code paths and share little beyond the
form of the basis functions; with $P=22$ and 130 decimal digits they agree to all 45 significant figures retained in all 84
cases underlying \tabl{tab:ions} (ten sectors, six ions, and both roots of each two-dimensional manifold), which bounds the error of each calculation, consistent with the convergence tests of \sect{sec:dirac}.

\paragraph{Level-resolved agreement.} The agreement extends to the individual terms of the spectral sum,
provided the terms are grouped by energy level. Consider the contribution to $E^{(2)}$ of the states of one
intermediate level $(n',j')$. In the Dirac theory it is the sum of
$|\langle\psi_\beta|\hat H_I|\bar\psi^{(0)}\rangle|^2/(E-E_{n'j'})$ over its states; in the \FV\ theory it is the corresponding sum of biorthogonal products $\langle\bar L|\hat H_I^\xi|\Psi_{\xi\beta}\rangle\langle L_\beta|\hat H_I^\xi|\bar\Psi_\xi\rangle/(E-E_{n'j'})$,
evaluated with the matrix elements~\eqref{eq:Ymatrix} extended to states of different levels. For all bound intermediate levels with $n'\le5$ that the field connects to any of the ten $(n,j,|m_j|)$ sectors of \tabl{tab:table2rev}, at $Z\alpha=0.05$, $0.3$ and $0.6715$ (234 comparisons, including all intra-shell partners), the two agree to at least 47 significant figures in 60-digit arithmetic. This is what the equivalence of the two theories leads one to expect: the equivalence does not depend on the field being static, so the dynamic polarizabilities of the two theories should coincide, and the level contributions are fixed by the residues of the dynamic polarizability at its poles. The comparison therefore tests the implementation of the
\FV\ formalism---the biorthogonal products, the $\vec\sigma\cdot\vec E$ term and the negative-norm partners---level by level rather than in the sum. The contributions of individual states do not agree: \tabl{tab:levels} shows, for the ground state of U$^{91+}$, that the \FV\ state $2s_{1/2}$ contributes $+0.0371$, whereas the Dirac $2s_{1/2}$ contributes nothing (by parity), and that the \FV\ $2p_{1/2}$ over-contributes by exactly the same amount. The \FV\ angular functions $\Theta_\xi=\Omega_\kappa+{\rm i}\bar\kappa\,\Omega_{-\kappa}$ mix parities, and a biorthogonal product, unlike a squared modulus, has no definite sign; the $\kappa$-labeled states are not parity eigenstates (\sect{sec:fvlevelshift}). State contributions are in any case not invariant within a degenerate level: they depend on the basis chosen in it, and only their sum over the level is basis independent. What the two theories share, for each pair of energy levels, is the level-to-level coupling of the static field: the products of the matrix elements of $\hat H_I$ (including, in the \FV\ theory, the $\vec\sigma\cdot\vec E$ term) summed over the states of both levels.

\begin{table*}[tbp]
\caption{Contributions of individual intermediate states $\beta$ and of whole levels $(n',j')$, for selected bound levels, to
$Z^4E^{(2)}/E_0^2$ for U$^{91+}$ ($Z\alpha=92/137$): Dirac theory, and \FV\ theory in the biorthogonal form of \eq{eq:fvDL}.
Upper block: the ground state; lower block: the lower root of the $n=2$, $j=\frac12$ manifold ($m_j=\frac12$). State
contributions depend on the basis chosen within a degenerate level and differ between the two theories; the level totals agree
to at least 47 significant figures (60-digit arithmetic).}
\label{tab:levels}
\begin{ruledtabular}
\begin{tabular}{llrr}
level $(n',j')$ & state $\beta$ & Dirac & \FV \\
\hline
$(2,\frac12)$ & $2s_{1/2}$ & $0$ & $+0.037\,141$ \\
 & $2p_{1/2}$ & $-0.286\,635$ & $-0.323\,776$ \\
 & level & $-0.286\,634\,747\,29$ & $-0.286\,634\,747\,29$ \\
$(2,\frac32)$ & $2p_{3/2}$ = level & $-0.541\,572\,791\,89$ & $-0.541\,572\,791\,89$ \\
$(3,\frac12)$ & $3s_{1/2}$ & $0$ & $+0.004\,508$ \\
 & $3p_{1/2}$ & $-0.029\,007$ & $-0.033\,515$ \\
 & level & $-0.029\,007\,140\,86$ & $-0.029\,007\,140\,86$ \\
$(3,\frac32)$ & $3p_{3/2}$ & $-0.085\,669$ & $-0.087\,125$ \\
 & $3d_{3/2}$ & $0$ & $+0.001\,455$ \\
 & level & $-0.085\,669\,443\,47$ & $-0.085\,669\,443\,47$ \\
\hline
$(1,\frac12)$ & $1s_{1/2}$ = level & $+0.143\,317\,373\,64$ & $+0.143\,317\,373\,64$ \\
$(3,\frac12)$ & $3s_{1/2}$ & $-0.475\,010$ & $+0.206\,565$ \\
 & $3p_{1/2}$ & $-4.860\,849$ & $-5.542\,423$ \\
 & level & $-5.335\,858\,203\,3$ & $-5.335\,858\,203\,3$ \\
$(3,\frac32)$ & $3p_{3/2}$ & $-4.112\,128$ & $-3.924\,431$ \\
 & $3d_{3/2}$ & $-15.162\,611$ & $-15.350\,308$ \\
 & level & $-19.274\,738\,901$ & $-19.274\,738\,901$ \\
\end{tabular}
\end{ruledtabular}
\end{table*}

\paragraph{Necessary ingredients and a computational advantage.} What makes the agreement informative is the list of
ingredients that had to work together: an eight-component solution space, an indefinite metric, a perturbation that differs
from the Dirac one, secular problems of mixed norm sign, biorthogonal left vectors, and negative-norm partners entering with
$\epsilon_k=-1$ (\sectsr{sec:fvdoubled}{sec:fvingredients}); for split levels the level shift must in addition carry its spin--gradient term
(\sect{sec:fvlevelshift}). The \FV\ formulation also has a practical merit, inherited from the second-order form of the theory.
The Dirac Dalgarno--Lewis operator is unbounded below and is inverted inside the spectral gap, where Galerkin approximations can
produce spurious solutions (spectral pollution~\cite{LewinSere2010}) unless the basis satisfies a balance condition~\cite{StantonHavriliak1984,Shabaev2004}. \sect{sec:dirac} shows that the usual (kinetic-balance) choice is not adequate here. The \FV\ equation~\eqref{eq:fvscalar} is of Schr\"odinger type and bounded below, since $\bar\gamma'(\bar\gamma'+1)=(\bar\gamma'+\frac12)^2-\frac14\ge-\frac14$ (in the $s$ channels, where $\bar\gamma'<0$, the basis $r^{\bar\gamma'+p}{\rm e}^{-\Lambda r}$ selects the Friedrichs extension); it is solved below its essential spectrum, where the min--max principle excludes spurious eigenvalues, and a single set of radial functions, without any balance condition,
converges exponentially in every channel.

\paragraph{What the agreement does not establish.} Since the exact \FV\ and Dirac Stark energies must coincide
(\sect{sec:intro}), the calculation cannot discriminate between the two theories: it validates the perturbative
formalism of the \FV\ theory and identifies the ingredients it requires. Nor is it a comparison with experiment:
within the Dirac model the relativistic linear Stark effect of hydrogen is confined to fields below about
$0.1$~kV\,cm$^{-1}$ (\tabl{tab:crossover}), where the shift is of order $10^{-8}$~a.u.\ and is dominated by the
effect of the Lamb shift. What experiment does use are the couplings $v_1$ and $v_3$ of the quench analyses (\sect{sec:fvlevelshift}); the relativistic reduction of $v_1^2$ by \eq{eq:f1closed}, $1.4\%$ at $Z=18$, was included in the analysis of the Ar$^{17+}$ measurement~\cite{HilleryMohr1980,GouldMarrus1983}. The results do bear on radiative transitions. In the long-wavelength limit and in the length gauge, an E1 rate is fixed by the transition energy and by the coupling of the two levels by a uniform electric field, the latter being the quantity compared above; if the \FV\ rate is defined consistently, through the biorthogonal products combined over the states of both levels, the present results imply that it equals the Dirac rate for the transitions examined here. Robson and Sutanto~\cite{RobsonSutanto2001} reported substantial differences between the eight-component and Dirac rates at large $Z$: for the $2P_{3/2}$--$1S_{1/2}$ transition of U$^{91+}$, $2.035\times10^{17}$ versus $1.580\times10^{17}$~s$^{-1}$. By contrast, the contribution of the $2p_{3/2}$ level to the static shift of the ground state is the same in the two theories to all figures computed ($-0.541\,572\,791\,89$ in units of $E_0^2/Z^4$, \tabl{tab:levels}). Their rates were obtained in the Coulomb gauge, with the full multipole expansion of the photon field, from the squared modulus of the $\tau_5$ matrix element of the interaction. For this transition each level is the only member of its manifold, so the mixing of $\kappa$-labeled states that accompanies split levels (\sect{sec:fvlevelshift}) does not arise, and in the length form the positive-norm coupling $|X+Y|$ and the biorthogonal one $2(XY)^{1/2}$ coincide with the Dirac value. Whether the reported difference originates in the velocity form of the coupling, in retardation or in the definition of the rate in that gauge is an open question that the formalism used here is suited to settle. A related consistency check exists for scattering: the eight-component theory reproduces the Klein--Nishina cross section for Compton scattering~\cite{RobsonSutanto2001b}.

\section{Conclusions and outlook}\label{sec:conclusions}

We have computed the first- and second-order Stark shifts, and their expansions in $Z \alpha$, of the $n=1$, $2$ and $3$ levels of six hydrogenic ions from H to U$^{91+}$ in the Schr\"odinger, Dirac and \FV\ theories, with all spectral sums evaluated exactly by the
Dalgarno--Lewis method. We have then compared the three descriptions level by level. Our conclusions are as follows.

\begin{enumerate}
\item \emph{The \FV\ formalism reproduces the Dirac Stark shifts through second order for $n\le3$.} The first-order roots coincide with the closed forms~\eqref{eq:f1closed} and the second-order shifts agree with the Dirac values to all 45 significant figures retained, for every level and every ion, with vanishing imaginary parts. The agreement also holds for
the contribution of every bound intermediate level tested ($n'\le5$), though not for individual states: the two theories share the level-to-level couplings of the field, while their state functions differ.

\item \emph{Both ingredients tested, the $\vec{\sigma} \cdot \vec{E}$ coupling and the negative-norm partners, are necessary.} Each degenerate \FV\ manifold contains states of both $\tau_5$ norms, so the reality of the \FV\ Stark spectrum is a result rather than a premise [\eq{eq:gensecular}]. Without the $\vec\sigma\cdot\vec E$ coupling every first-order root acquires an imaginary part of order $\alpha E_0$; without the negative-norm partners the roots are real but have the wrong $(Z\alpha)^2$ coefficient.

\item \emph{Split levels need the same care.} Norm sign and parity are locked in the doubled space [\eq{eq:fvparity}], so a change of the electrostatic potential that lifts the $2s_{1/2}$--$2p_{1/2}$ degeneracy must enter through its \FV\ image, spin--gradient term included [\eq{eq:fvdV}]; the resulting operator is then diagonal in fixed combinations $t_s$, $t_p$ of the $\kappa$-labeled states [\eq{eq:cmix}], and the \FV\ theory supplies the Dirac couplings $v_1$ and $v_3$ on which field-quenching and anisotropy measurements of the Lamb shift rely. If placed by hand on the $\kappa$-labeled states, the level shift would produce spurious imaginary linear Stark term $\pm\frac32{\rm i}\alpha E_0$. The closed form~\eqref{eq:f1closed} of $v_1$ reproduces the numerical results of Hillery and Mohr~\cite{HilleryMohr1980}.

\item \emph{The relativistic and nonrelativistic linear effects differ, but the linear strength is conserved.} The relativistic roots are smaller by finite factors ($\sqrt3$ at $n=2$) and form a different pattern, because relativity replaces the Runge--Lenz degeneracy of the shell with the degeneracy of $\pm\kappa$; the missing strength is carried by the singular $(Z\alpha)^{-2}$ terms of the second order, as the sum rules~\eqref{eq:sumrule1} and~\eqref{eq:secondmoment} show. The two regimes are the limits of a nearly universal crossover in $E_0/(Z\Delta_2)$, whose midpoint lies near $2.8$~kV\,cm$^{-1}$ for hydrogen within the Dirac model and beyond the field-ionization threshold for U$^{91+}$.

\item \emph{The second-order shifts have a transparent analytic structure.} For levels without a fine-structure partner the relativistic correction stiffens the atom, by $44\%$ for the ground state of U$^{91+}$ but only $13\%$ and $6\%$ for the nodeless $2p_{3/2}$ and $3d_{5/2}$ states. For the ground state, our values agree with the closed form of Szmytkowski~\cite{Szmytkowski1997} to 45 significant figures. For the other levels the shift separates exactly into an intra-shell part, known in closed form and containing all singular terms, and a regular part; the decomposition derives the $(Z\alpha)^0$ coefficient of these levels analytically, places the $\pi^2$ terms of the $(Z\alpha)^2$ coefficients in the intra-shell part (the remainder being rational to the 26 figures of the identification), and gives their coefficients in terms of the singular ones [\eq{eq:bformula}].

\item \emph{The continuum cannot be omitted.} Nonrelativistically, the discrete spectrum supplies $81.4\%$ of the ground-state polarizability and between $97.6\%$ and $99.1\%$ of the $n=3$ shifts, so that a bound-state-only calculation would be wrong at the $19\%$ level at $n=1$.

\item \emph{The perturbative window set by the fine structure widens as $Z^5$} (\tabl{tab:validity}); for the heaviest ions
field ionization is the limit. Numerically, the Dirac Galerkin basis must respect the small-$r$ behavior of both components
(\sect{sec:dirac}), whereas the scalar \FV\ equation needs no balance condition.
\end{enumerate}

Taken together, these results answer the question posed in the introduction. Within the scope tested---first and second order, $n\le3$, $Z\le92$---the \FV\ theory gives a consistent description of a spin-$\frac12$ particle in a static electric field, assuming the completeness of its biorthogonal set of states: its doubled solution space, indefinite metric and biorthogonal structure are the elements through which it reproduces the Dirac response, order by order and level by level. Where the two theories may still appear to differ is in quantities whose definition goes beyond the levels and their couplings. Robson and Sutanto~\cite{RobsonSutanto2001} reported radiative transition probabilities that differ from the Dirac ones; the present results imply that, in the long-wavelength limit of the length gauge and with the rates defined through biorthogonal products, they coincide for the transitions examined here, so that the origin of the reported difference remains to be identified. Closed-form first-order roots, $Z\alpha$ expansions of all second-order shifts through $(Z\alpha)^2$ (and $(Z\alpha)^4$ for the levels without a partner) and the sum rules connecting them to the nonrelativistic results are provided as benchmarks. Radiative (QED) corrections, the finite nuclear size, and the nuclear recoil are not included in the Stark shifts computed here; model level shifts serve to show how such corrections must enter the \FV\ formalism.

A natural next step is the Zeeman effect. A uniform magnetic field is even under parity, and in \eq{eq:fvH} the magnetic spin term enters both blocks with the same sign and a real coefficient, unlike the electric term. The exact answer is again fixed---for the ground state the Breit value $g_{1s}=\frac23\bigl[1+2\sqrt{1-(Z\alpha)^2}\bigr]$~\cite{Breit1928}---so the Zeeman calculation would test a different part of the same formalism (the minimal-coupling term of $\mathbf D^2$ and the mixed-norm secular problem for a time-odd perturbation) in a system whose bound-electron $g$ factor is measured with extraordinary precision~\cite{Sailer2022,Morgner2023}. A second natural step is to compute \FV\ radiative transition rates in the length and velocity gauges, with and without retardation, and so settle the question raised in \sect{sec:DvsFV}.

\begin{acknowledgments}
We are grateful to Dr. Haryanto Mangaratua Siahaan for pointing out that the perturbing Hamiltonian cannot simply be inserted into the \FV\ Hamiltonian in the way it is in the Schr\"{o}dinger and Dirac theories, a remark that shaped the treatment of the field in Section~\ref{sec:fv}. We also thank Ms. Jessica Belicia Cahyono for her assistance in constructing the initial version of the Mathematica code from which the numerical results of this work grew. This work was supported by Parahyangan Catholic University under research grant contract No. III/LPPM/2025-02/49-P.

We used Claude (Opus 5, Anthropic) as an AI assistant in the preparation of this work. It was used to check and debug the Mathematica code built on the initial version mentioned above, which evaluates the Stark shifts reported in Tables~\ref{tab:schr}--\ref{tab:ingredients}. The authors formulated the problem and the theoretical approach, directed each task, and verified all AI-generated outputs.  The code was run independently by the authors in Mathematica~15 and its results were checked against the closed-form expressions derived in this work. The results were also checked against the nonrelativistic parabolic formula~\cite{Bethe1957} and the results of Hillery and Mohr~\cite{HilleryMohr1980}, and every derivation, statement and reference in the text was checked by the authors. The authors take full responsibility for the content of this article.
\end{acknowledgments}

\section*{Data availability}
The Mathematica code that support the findings of this article cannot be made publicly
available because it is part of ongoing research by the authors. Request for the notebook should be sent to the authors.

\bibliography{stark}

\end{document}